\documentclass[11pt,a4paper]{article}

\usepackage{jcappub}
\usepackage[utf8]{inputenc}

\usepackage[cm]{fullpage}
\usepackage{siunitx}

\newcommand{\beq}{\begin{equation}}
\newcommand{\eeq}{\end{equation}\\}
\newcommand{\beqa}{\begin{eqnarray}}
\newcommand{\eeqa}{\end{eqnarray}\\}

\newcommand{\doverc}{\frac{D}{C}}
\newcommand{\al}[1]{\alpha_{#1}}
\newcommand{\be}[1]{\beta_{#1}}
\newcommand{\alb}[1]{\bar{\alpha}_{#1}}
\newcommand{\beb}[1]{\bar{\beta}_{#1}}
\newcommand{\ga}{\gamma}
\newcommand{\rhochi}{\rho_\chi}
\newcommand{\pchi}{p_\chi}
\newcommand{\cpoc}{\frac{C'}{C}}
\newcommand{\dpod}{\frac{D'}{D}}
\newcommand{\cppoc}{\frac{C''}{C}}
\newcommand{\dppod}{\frac{D''}{D}}
\newcommand{\gbp}[2]{\left(#1\ga^2 + #2\right)} 
\newcommand{\gbm}[2]{\left(#1\ga^2 - #2\right)} 

\newcommand{\phidot}{\dot{\phi}}
\newcommand{\chidot}{\dot{\chi}}
\newcommand{\phidotdot}{\ddot{\phi}}
\newcommand{\chidotdot}{\ddot{\chi}}

\DeclareSIUnit\parsec{pc}
\newcommand{\bd}{{\rm d}}

\newcommand{\pmat}[1]{\begin{pmatrix}#1 \end{pmatrix}}

\begin{document}

\hspace{5.2in} \mbox{NORDITA-2015-121}\\\vspace{1.53cm} 

\title{Disformally coupled inflation}

\author[a]{Carsten van de Bruck}
\author[b]{, Tomi Koivisto}
\author[a]{and Chris Longden}

\affiliation[a]{Consortium for Fundamental Physics, School of Mathematics and Statistics, University of Sheffield, Hounsfield Road, Sheffield S3 7RH, United Kingdom}
\affiliation[b]{Nordita, KTH Royal Institute of Technology and Stockholm University, 
Roslagstullsbacken 23, SE-10691 Stockholm, Sweden}

\abstract{A disformal coupling between two scalar fields is considered in the context of cosmological inflation. The coupling introduces novel derivative interactions mixing the kinetic terms of the fields but without introducing superluminal or unstable propagation of the two scalar fluctuation modes. Though the typical effect of the disformal coupling is to inhibit one of the fields from inflating the universe, the energy density of the other field can drive viable near Sitter -inflation in the presence of nontrivial disformal dynamics, in particular when one assumes exponential instead of power-law form for the couplings. The linear perturbation equations are written for the two-field system, its canonical degrees of freedom are quantised, their spectra are derived and the inflationary predictions are reported for numerically solved exponential models. A generic prediction is low tensor-to-scalar ratio.}

\maketitle

\section{Introduction} \label{Sec:Intro}

The observations of the anisotropies in the cosmic microwave background (CMB) due to the WMAP \cite{Larson:2011rwv} and Planck  \cite{Ade:2013zuv,Ade:2015lrj} missions in recent years have given us a high precision testing ground for theories of the physics of the early universe. In particular, the inflationary paradigm's generic prediction of a nearly scale-invariant power spectrum of primordial fluctuations is widely consistent with the data, cementing its role in our modern understanding of cosmology. While inflationary theory is successful as a general paradigm, the microphysics of the early universe remain unknown and as such, the exact mechanism responsible for inflation is still unclear. Modifications of gravity, as well as the effects of extra dimensions and/or multiple scalar fields inspired by string theory or supersymmetry may play an important role. As more experimental data is accumulated and its analysis constrains more and more parameters that are influenced by the mechanism driving inflation, such as the tensor-to-scalar ratio and non-gaussianity, the number of feasible theories will continue to shrink. Of particular interest at present, the continuing lack of a positive detection of primordial gravitational waves has constrained the amplitude of tensor fluctuations during inflation to be in considerable tension with the predictions of the simplest models of inflation. \\

Usually the potential of the inflaton scalar field should be extremely flat for its nearly constant energy density to generate a nearly de Sitter expansion of the spacetime it fills. This is challenging to realise without fine-tunings in high energy theories, and there is no compelling embedding of inflation to string theory though many interesting constructions have been elaborated in the literature \cite{Baumann:2014nda}. In the Dirac-Born-Infeld or DBI-inflation, the flatness problem is bypassed by noncanonical kinetic terms, which can lead to effectively slowly rolling dynamics in steep potential slopes \cite{Silverstein:2003hf}. In DBI inflation, the scalar fields describing the location of a D-brane in a higher dimensional warped geometry from the four-dimensional point of view assume kinetic terms of a suitable form such that the effective kinetic contribution to the energy density is screened by the warping. Furthermore such nontrivial dynamics can result in observable non-Gaussianities of a distinctive form \cite{Alishahiha:2004eh}. Though in the simplest case one has just one D3-brane moving in a radial direction of the warped geometry, one can easily imagine (though less easily explicitly construct) various more complicated set-ups, for example when the brane is dynamical also in the angular directions, which would then be described at the effective four-dimensional level as multi-field inflationary models with non-canonical, coupled kinetic terms. General and model-independent formalisms have been developed to study such multi-field scenarios in particular by Langlois {\it et al} \cite{Langlois:2008mn,Langlois:2008qf}, who allowed an arbitrary number of scalar fields with a nonlinear sigma-type quadratic kinetic matrix. In general, myriads of single- and multi-field scalar models with more and less convincing pretensions for theoretical motivations have been proposed, with some examples listed in the recent encyclopaedic compendiums of references \cite{Martin:2013tda} and \cite{Vennin:2015vfa}.  \\

In this article, we will focus on non-conformal or {\it disformal} couplings, a presently popular tool in cosmological model building that has however been largely neglected in the context of inflation. Nevertheless, disformal couplings naturally arise also in the DBI-type scenarios, since there the induced metric upon a moving brane is disformal. It is this, in the first place, that the non-canonical (coupled) kinetic terms for the field(s) describing the extra-dimensional coordinate(s) of the brane stem from. Furthermore, it would result in a disformal coupling of any other field(s), fundamental or effective, residing upon such a moving brane \cite{KOIVISTO:2013jwa}. From the very general viewpoint of an arbitrary viable scalar-tensor theory, significant classes of Horndeski actions can be equivalently written as Einstein's theory with a disformally coupled matter sector \cite{Zumalacarregui:2012us,Bettoni:2013diz,Zumalacarregui:2013pma}. \\

The disformal equivalence of theories has been recently studied in detail specifically in the context of cosmological perturbations  \cite{Minamitsuji:2014waa,Domenech:2015hka,Tsujikawa:2015upa,Motohashi:2015pra} which is also interesting for the inflationary generation of cosmological structure. To date we however have at hand neither specific viable models of disformalised inflation nor detailed understanding of their generic features. Kaloper \cite{Kaloper:2003yf} proposed a model with a disformal coupling of a massless inflaton to the Standard model, and found that such coupling could be well constrained by collider experiments such as the LHC, as confirmed by more recent studies \cite{Brax:2014vva,Brax:2015hma}, though these constraints can be avoided if the disformal coupling is suppressed in the Standard model sector after reheating. On the other hand, in the context of late-time cosmology and dark energy, aspects of disformal couplings such as the possible disformal screening mechanism \cite{Koivisto:2012za,Bettoni:2015wta} and its limitations \cite{Sakstein:2014isa,Ip:2015qsa}, background dynamics \cite{Sakstein:2014aca,vandeBruck:2015ida,Sakstein:2015jca}, self-couplings  \cite{Koivisto:2008ak,Zumalacarregui:2010wj}, DIMP dark matter \cite{KOIVISTO:2013jwa,Koivisto:2013fta}, structure formation \cite{Bettoni:2012xv,Hagala:2015paa,Gleyzes:2015pma}, photon interactions \cite{vandeBruck:2013yxa,Brax:2013nsa,vandeBruck:2015rma} and astrophysical implications \cite{Koyama:2015oma,Koivisto:2015mwa}, have recently been investigated, see e.g. \cite{Sakstein:2015oqa,Vu:2015jja} for introductory treatments.  Most of these studies take the disformally coupled matter to be modelled by the perfect fluid parameterisation. However, the workings of nonminimal couplings depend upon the precise form of the matter fields and therefore the disformalisation of an effective perfect fluid parameterisation (of a fundamental field) results in general in a different and thus less correct theory than that of a disformalised fundamental field (in the effective perfect fluid parameterisation) though in the minimally coupled case the descriptions are equivalent \cite{Koivisto:2015qua}.  \\

Here we will analyse perhaps the simplest example of a nontrivial disformal field theory: a canonical scalar field coupled to the disformal metric defined by another canonical scalar field. This defines a two-field system that we will apply to model cosmological inflation. The article will proceed as follows. In Section \ref{Sec:Model} we will first present our action together with its extra dimensional motivations, and its equations of motion which we then adapt to cosmology, with some of the more technical details of the derivations in the section confined to the three appendices \ref{App:XYZ}-\ref{App:AlphaBetaBar}. In Section \ref{Sec:Numerics} we then analyse the system of cosmological perturbations. In particular, we extract the properties of the two new propagating degrees of freedom in cosmological backgrounds, and discuss the issue of setting the initial conditions for inflation.  We are then ready to run numerical tests to obtain the inflationary dynamics in the various different cases it turns out we can construct with the disformally coupled two-field model. Section \ref{Sec:ResultsBG} is devoted to numerical study of the models. We conclude in Section \ref{Sec:Conclusions} with an outlook to further research.

\section{The model} \label{Sec:Model}

We begin by presenting our coupled two-field two-field action and the corresponding equations of motion in their covariant form below in \ref{Sec:Action}, discuss their possible derivation in brane world models (\ref{Sec:String}) and specify the couplings to be used in the following calculations with more phenomenological aims (\ref{Sec:Phenom}). In part \ref{Sec:Background} we then adapt the equations for cosmology, writing down, as conventional, first the Friedmann equations for the homogenenous background and then the field equations for cosmological perturbations. 

\subsection{Action and equations of motion} \label{Sec:Action}

We want to get set out to explore a gravitating system where the novel ingredient is disformal coupling of scalar field to another. For facilitate this exploration, we would like to choose the ingredients of the model in a minimalistic way. Therefore, let us from the beginning assume that there exists a frame wherein it holds that 1) the pure gravity sector is GR 2) the disformal metric is given by only one field that furthermore has a canonical lagrangian  3) also the disformally coupled field is canonical. \\

The full action for this theory is thus fixed as the following (in Planck units):
\begin{align}
S= \frac{1}{2} \int {\rm d}^4 x \sqrt{-g} \ R \ - \ \int {\rm d}^4 x \sqrt{-g}\left[\frac{1}{2} g^{\mu\nu}\phi_{,\mu}\phi_{,\nu} + U(\phi)\right] \ - \ \int {\rm d}^4 x \sqrt{-\hat{g}}\left[ \frac{1}{ 2}\hat{g}^{\mu\nu}\chi_{,\mu}\chi_{,\nu} + V(\chi)\right] \,. \label{eq:DisformalAction}
\end{align}
The model can be then seen as a very generic scalar-tensor theory described by the action $S_{ST}(g,\phi)$ (in the Einstein frame), coupled to matter fields $\chi$ that propagate not on the gravitational metric $g$, but on a second metric (such as an induced metric on a brane) denoted by $\hat{g}$ as $S  = S_{ST}(g, \phi) + S_{M}(\hat{g}, \chi)$. The relation between the two metrics is in general given by the disformal transformation, defined by
\beq \label{eq:DisformalMetric}
\hat{g}_{\mu \nu} = C(\phi) g_{\mu \nu} + D(\phi) \phi_{,\mu} \phi_{,\nu} \,,
\eeq
where $C(\phi)$ is the conformal coupling function, and $D(\phi)$ is the disformal coupling function. Both $C$ and $D$ are functions of a scalar field $\phi$. In more general cases, physically consistent relations could depend also upon the kinetic term $X = g^{\mu\nu}\phi_{,\mu}\phi_{,\nu}$, and furthermore the relation could in principle involve both of the scalar fields and their derivatives -- but once again, in this exploratory study our approach is minimalistic rather than aiming at great generality, and this fixes our disformal relation to the relatively simple form of (\ref{eq:DisformalMetric}). \\

We have hence specified how the metric for the $\chi$-matter in the action for this theory is disformally related to the gravitational metric. Using the relation (\ref{eq:DisformalMetric}) to rewrite the action in terms of the latter metric alone we obtain:
\beq \label{eq:Action2}
S = \frac{1}{2}\int {\rm d}^4 x \sqrt{-g}\left[ R - g^{\mu\nu}\phi_{,\mu}\phi_{,\nu} - 2U - \frac{C}{\ga}\left( g^{\mu\nu}\chi_{,\mu}\chi_{,\nu} + 2 C V\right) + \ga D \left(\phi_{,\sigma}\chi^{,\sigma}\right)^2 \right] \,,
\eeq
where we have suppressed the arguments of $C(\phi)$, $D(\phi)$, $U(\phi)$ and $V(\chi)$ for brevity, and for convenience defined the parameter $\ga$ as
\beq
\label{eq:gamma_general}
\ga = \left(1 + \frac{D}{C} g^{\mu\nu}\phi_{,\mu}\phi_{,\nu}\right)^{-\frac{1}{2}} \,.
\eeq
In making the above substitution, we have hence gone from an action describing two scalar fields with no explicit interactions and on different metrics, to a theory of two scalar fields with explicit interaction terms but living on the same metric. That is, we have turned $S_{M}(\hat{g}, \chi)$ into $S'_{M}(g, \phi, \chi)$. \\

The field equations follow by varying  (\ref{eq:Action2}) with respect the metric $g_{\mu\nu}$ as
\beq
G_{\mu\nu} = T_{\mu\nu} = T^{(\phi)}_{\mu\nu} + T^{(\chi)}_{\mu\nu} \,,
\eeq
where $T^{(\phi)}_{\mu\nu}$ is the usual energy-momentum tensor for a minimally-coupled scalar field but $T^{(\chi)}_{\mu\nu}$ assumes interesting cross-terms due to the coupling:
\beq
T^{(\phi)}_{\mu\nu} = -\left( \frac{1}{2} g^{\alpha\beta}\phi_{,\alpha}\phi_{,\beta} + U \right)g_{\mu\nu} + \phi_{,\mu}\phi_{,\nu}\,,
\eeq
\beqa
\label{eq:EMTchi}
 T^{(\chi)}_{\mu\nu} =   & - & \left[\frac{C}{\ga}\left( \frac{1}{2} g^{\alpha\beta}\chi_{,\alpha}\chi_{,\beta} + CV\right)-\frac{1}{2}\ga D \left(\phi_{,\sigma}\chi^{,\sigma}\right)^2\right] g_{\mu\nu} +\frac{C}{\ga} \chi_{,\mu}\chi_{,\nu}  - 2\ga D\left(\phi_{,\sigma}\chi^{,\sigma}\right) \chi_{,(\mu}\phi_{,\nu)} \nonumber \\
& + & \left[ \ga D \left( \frac{1}{2} g^{\alpha\beta}\chi_{,\alpha}\chi_{,\beta}+ CV\right) + \frac{\ga^3 D^2}{2C} \left(\phi_{,\sigma}\chi^{,\sigma}\right)^2\right] \phi_{,\mu}\phi_{,\nu} \,.
\eeqa
Varying the action with respect to $\chi$ we obtain the equation of motion,
\beqa
\label{eq:gen_KGchi}
\Box \chi - CV' & - & \frac{\ga^2}{2}\left[ \left( \ga^2-3 \right) \cpoc - \left( \ga^2-1\right)  \dpod \right] \left(\phi_{,\sigma}\chi^{,\sigma}\right) \nonumber \\
& - & \ga^2\frac{D}{C}\left[ \phi^{,\mu}\phi^{,\nu}\nabla_\mu\chi_{,\nu} +  \left(\phi_{,\sigma}\chi^{,\sigma}\right) \Box\phi\right] + \ga^4 \frac{D^2}{C^2} \left(\phi_{,\sigma}\chi^{,\sigma}\right) \phi^{,\mu}\phi^{,\nu}\nabla_\mu\phi_{,\nu} = 0 \,,
\eeqa
where priming denotes the derivative with respect to $\phi$, except, of course, in the case of $V'$ where it denotes the derivative with respect to $\chi$. \\

We also obtain an equation of motion for $\phi$ by noting that conservation of the energy-momentum tensor implies $\nabla^\mu T_{\mu\nu} = \nabla^\mu \left(T^{(\phi)}_{\mu\nu} + T^{(\chi)}_{\mu\nu}\right) = 0$, and hence,
\beqa
\nabla^\mu T^{(\phi)}_{\mu\nu} = Q \phi_{,\nu} \,, \quad
\nabla^\mu T^{(\chi)}_{\mu\nu}  = - Q \phi_{,\nu} \,,
\eeqa
where we have defined the interaction term $Q$, that, by using the second equality, turns out as
\beq
\label{eq:Q}
Q = \ga^2 \left[T^{(\chi)\mu\nu} \nabla_\mu \left(\doverc \phi_{,\nu}\right)-\frac{1}{2C}\left( C' T^{(\chi)} + D' T^{(\chi) \mu\nu}\phi_{,\mu}\phi_{,\nu}\right)\right] \,,
\eeq
where $T^{(\chi)}$ is the trace of $T^{(\chi)}_{\mu\nu}$. Hence we find the equation of motion for $\phi$ to be
\beq
\label{eq:gen_KGphi}
\Box \phi - U' - Q = 0 \,,
\eeq
with $Q$ given by the rather complicated formula above.

\subsubsection{Realisation in brane and string models} \label{Sec:String}

The two-field model (\ref{eq:DisformalAction}) represents a general scalar-tensor theory with the matter sector given by a single scalar field. \\

From a phenomenological point of view, the coupling (\ref{eq:DisformalMetric}) can be seen as a generalisation, or maybe rather a variation, of multi-field inflation models on which there exists an extensive literature. General formalisms have been developed to tackle systems with multi-field models,
which however often technically rely upon the quadratic restriction for the nonlinear sigma metric, which would require a Finslerian extension to incorporate our set-up. One main motivation to study the quadratic models in particular were the Dirac-Born-Infeld (DBI) brane scenarios \cite{Silverstein:2003hf,Alishahiha:2004eh}, which can indeed have effective four dimensional descriptions wherein the fields mutually combine into a quadratic kinetic term whose total action then assumes the DBI form. However, the couplings of our model do not fall into this class due to the peculiar form of the derivative interactions generated by (\ref{eq:DisformalMetric}). Nevertheless, as we will argue below, action (\ref{eq:DisformalAction}) can actually also be well motivated in DBI-type brane scenarios. \\

Basically this requires simply that not all the branes are empty: the matter fields upon a moving brane will then be disformally coupled in the effective four-dimensional geometry. In the ''dark D-brane world'' scenarios the brane energy accelerates the universe and dark matter is made of the disformally interacting massive particles (DIMPs) on the brane \cite{KOIVISTO:2013jwa,Koivisto:2013fta}. In previous studies of this and other late-time disformal cosmologies, the coupled (dark or otherwise ) matter has been usually parameterised as a perfect fluid with a constant equation of state. It is easy to see that such a description is too simple to capture the full structure of the underlying disformally coupled field theory even in the case of a scalar field (though it can be of course that in some regions it provides an adequate approximation). It is thus of some interest on its own to consider such field theories, and to our knowledge what will be presented in the following are some first basic results for the case that seems the simplest and quite natural starting point, a scalar field living in a disformal metric (given by another field). Note though that recently cosmology was studied with a vector field living on a brane, i.e. a (massive) vector coupled to a disformal metric given by a DBI type scalar field \cite{Koivisto:2014gia,Koivisto:2015vda}. \\

To briefly sketch the brane world picture we will, for concreteness, consider a 3-brane in a 10-dimensional spacetime with the extra dimensions compactified. Assigning the coordinates $x^\mu$ to the noncompact dimensions, where $\mu=(0,...,3)$, and the coordinates $y^A$ to the compact dimensions, with $A=(4,...,9)$, the ten dimensional metric  takes the form
\beq \label{GMNintro}
G_{MN}{\rm d}x^M {\rm d}x^N = h^{-1/2}(y^A)g_{\mu\nu}{\rm d}x^\mu {\rm d}x^\nu + h^{1/2}(y^A)g_{AB}{\rm d}y^A {\rm d}y^B\,,
\eeq
where $g_{AB}$ is the metric of the internal six dimensional Calabi-Yau manifold, and in order to preserve Lorentz symmetry in the noncompact dimensions, the warp factor $h$ is a function of only the internal coordinates, $h=h(y^A)$.  Defining the coordinates $\xi^a$ on the world-volumes of the 3-brane, where $a=(0,...,3)$, we can embed it into the spacetime by the mapping $x^M(\xi^a)$. In the noncompact dimensions, we choose the static gauge $\xi^\mu = x^\mu$, whereas in the compact dimensions the embedding functions are kept general, $y^A=y^A(\xi^a)$ to allow the brane to float around dynamically. For simplicity, we assume that the brane is moving only in a radial direction $r$. The induced metric upon the brane then becomes
\beq\label{inducedmetric}
\bar g_{\mu\nu} = G_{MN}\partial_\mu x^M \partial_\nu x^N = h^{-1/2}(r) g_{\mu\nu}+h^{1/2}(r)\partial_\mu r\partial_\nu r\,.
\eeq
Here we can already identify the form of the couplings in terms of the warp factor  $C(r)^{-1} = D(r) = h(r)^{1/2}$ and the scalar field with the brane's position $\phi \propto r$. Our physical metric in this picture is the bulk spacetime metric $g_{\mu\nu}$, while the scalar field $\xi$ that we take to live upon the moving 3-brane, is minimally coupled to the induced metric (\ref{inducedmetric}) that naturally has the disformal relation (\ref{eq:DisformalMetric}) to $g_{\mu\nu}$ with $CD=1$ that stems from the geometrical construction (\ref{GMNintro}).  \\

Let us consider such constructions in a bit more detail. The brane is associated with the tension $T_3$, that has the dimension of energy density. The canonically normalised radial coordinate is $\phi\equiv \sqrt{T_3}r$ and the corresponding warp factor is $h(\phi)\equiv T_3^{-1}h(r)$, for the radial direction  $r$ in the warped geometry. In the DBI scenario, the scalar action for a D3-brane would assume the form\footnote{This simply stems from the determinant of the brane metric when one adds a charge term and takes into account a potential for the field.}
\beq\label{Sphi}
S_{\phi} = -\int {\rm d}^4 x \sqrt{-g} \left[ h^{-1}(\phi)\!\left( \sqrt{1+h(\phi)\partial_{\mu}\phi \partial^{\mu}\phi}-1\right) + V(\phi)\right]\,.
\eeq
When the brane is moving slowly enough and the warp is not too strong, this is well approximated by the canonical lagrangian we employ in actual calculations in this paper. Once we consider the scalar field $\chi$ to live upon the brane however, its action becomes precisely as stated in (\ref{eq:DisformalAction}),  with the conformal and disformal factors $C(\phi) \equiv (T_3 \,h(\phi))^{-1/2}$ and $D(\phi) \equiv (h(\phi)/T_3)^{1/2}$, respectively. Given an extra-dimensional geometry to determine the warp factor $h$ and the details of the compactification set-up to determine the potentials $V(\phi)$ and $U(\chi)$, we can thus in principle derive the precise form and parameters of our model from first principles. \\

In practise it is not an easy task to construct such set-ups in robust detail \cite{Baumann:2010sx}, and only a few explicit computations have been carried out, notably in the ''warped throat'' -type geometries where the bulk spacetime becomes strongly warped at some region (''the throat'', corresponding to $r\rightarrow 0$) due to presence energy of sources at $r=0$.
The prototype anti de Sitter (AdS) geometry, which is smooth all the way to the tip of the throat, is given by the compact version of the more complete Klebanov-Strassler geometry that is an exact supergravity solution \cite{Klebanov:2000hb}. The energy sources in that case are fluxes due to wrapped D3 and D5-branes. The full Klebanov-Strassler geometry is described a more complicated warp factor $h$, but however in an interior region far enough from the throat it may be approximated by the simpler AdS$_5\times$ S$^5$ geometry, which corresponds to the near horizon limit of a stack of $N$ D3-branes \cite{Giddings:2001yu}. The warp factor in this region is given by the quartic power of the inverse distance as\footnote{For the Klebanov-Srassler geometry, this AdS$_5$ approximation breaks down near the tip of the throat. There the warp factor approaches a constant value  $h\rightarrow$ const.$({\cal O}(1))$ with corrections of order $ {\cal O}(r^2)$.}
$h=\frac{\lambda_{AdS}}{r^4}$, with the t'Hooft coupling, $\lambda_{AdS}$ given by $\lambda_{AdS} = 4\pi\alpha'^2 g_s N$, where  $\alpha'$ is the string coupling, $g_s N \gg1$ for the supergravity approximation to be valid, while $g_s<1$ for string perturbation theory approximation to hold, so that the t'Hooft coupling, $\lambda_{AdS} \gg 1$. With the identification $\lambda_{AdS}=1/T_3$ we thus arrive at:
\beqa
C(\phi) = \frac{\phi^2}{\sqrt{T_3}}\,, \label{eq:StringC}\\
D(\phi) = \frac{1}{\sqrt{T_3}\phi^{2}}\,. \label{eq:StringD} 
\eeqa
Implementing these couplings in the action eq. (\ref{eq:DisformalAction}) thus represents a ''stringy'' realisation of our inflationary model. \\

In the following we shall however also take some liberties in exploring different forms of the couplings (in addition to considering the canonical instead of the DBI form for the field $\phi$). Though only one example of exact solution for warped geometry is available in the literature (approximated by ({\ref{eq:StringC},\ref{eq:StringD})), there is no reason why many kinds of consistent solutions for different shapes of geometries would not exist. Such would then result in different functional forms for the couplings $C$ and $D=1/C$. The main point we wanted to make here is simply the disformal coupling (\ref{eq:DisformalAction}) emerges generically in the effective description of extra dimensional geometries with the generic warped form (\ref{GMNintro}).


\subsubsection{Phenomenological couplings} \label{Sec:Phenom}

Aside from the string theory motivated couplings discussed in section  \ref{Sec:String} we will also consider models where the couplings are given by
\beqa
C(\phi) = C_0 e^{c \phi} \label{eq:PhenomC} \, , \\
D(\phi) = D_0 e^{d \phi} \label{eq:PhenomD} \, .
\eeqa
Here, we use four parameters, $C_0$, $c$, $D_0$ and $d$ to describe our coupling functions, enabling us to observe a wider range of behaviours. An interesting special case of this parametrisation is when $c = d = 0$ and the couplings become constants,
\beqa
C(\phi) = C_0 \, , \\
D(\phi) = D_0 \, .
\eeqa
This limit describes the brany scenarios described in previous subsection \ref{Sec:String} in the limit of trivial extra dimensional geometry, i.e. when the warp $h$ of the induced metric  (\ref{inducedmetric}) is a constant. In this setting the disformal interaction is hence more generic than the conformal, as the latter vanishes with the gradients of the warp. Then the scale $D_0=T_3^{-1}$ can be identified with the (inverse) tension of the 3-brane $T_3$.  We can also associate a  ``disformal mass'' scale $m_D$ to this, which is then, since $D$ has a mass dimension of $-4$, 
\beq
m_D = D_0^{-\frac{1}{4}} = T_3^{\frac{1}{4}}\,.
\eeq
The disformal mass scale is a convenient descriptor of the magnitude of the disformal coupling.

\subsection{Cosmological equations} \label{Sec:Cosmology}

To study the cosmological implications of disformal couplings, we wish to take the equations of motion found in section \ref{Sec:Action} and choose $g$ to be a spatially-flat Friedmann-Robertson-Walker metric,
\beq \label{frw}
\bd s^2 = -\bd t^2 + a^2(t)\left( \bd x^2 + \bd y^2 + \bd z^2 \right)\,.
\eeq
Furthermore we assume the fields are homogeneous to a good approximation, so we can describe their spatial dependence as small perturbations on a background.

\subsubsection{Background} \label{Sec:Background}

At the background level, we first note that $\ga$ takes the form
\beq
\label{eq:gamma}
\ga = \left(1 - \frac{D}{C} \dot{\phi}^2\right)^{-\frac{1}{2}} \, ,
\eeq
where the dot represents differentiation with respect to cosmic time, $t$.  We also note from this expression that $\ga \in \left[1,\infty \right)$ for positive coupling functions $D$ and $C$. $\ga$ can instead take values less than unity for a negative choice of $D$, but as we will see later in section \ref{Sec:BG_Phenom} this does not lead to physically desirable outcomes and as such this case is ignored.\\ 

Using the cosmological metric and homogeneous fields, we find the Friedmann for the background to be,
\beqa
3H^2 & = & \rho  \ =  \ \rho_\phi + \rhochi \ = \  \frac{1}{2}\dot{\phi}^2 + U + \rhochi  \, ,\label{eq:EE1} 
\\
-\left(2\dot H+3H^2\right) & = &  p \ = \ p_\phi + \pchi  \ = \ \frac{1}{2}\dot{\phi}^2 - U +  \pchi \, , \label{eq:EE2}
\eeqa
where $H \equiv \dot{a}/a$ is the Hubble parameter, and the effective energy density $\rhochi$ and pressure $\pchi$ of the $\chi$ field are found to be,
\beqa
\rhochi = \gamma C\left( \frac{1}{2}\gamma^2\dot{\chi}^2+CV\right) \, ,\label{eq:rhochi} 
\\
\pchi = \frac{C}{\gamma} \left( \frac{1}{2}\gamma^2\dot{\chi}^2 - CV\right) \, . \label{eq:pchi}
\eeqa
We also specialise the Klein-Gordon equations for $\phi$ and $\chi$, eqs. (\ref{eq:gen_KGchi}--\ref{eq:gen_KGphi}), to a cosmological background and find that, 
\beq
\label{eq:KGchi} 
\ddot\chi +3H\dot\chi + \gamma^2 \doverc \dot{\phi}\ddot{\phi}\dot{\chi} +\gamma^{-2}C V' = \frac{1}{2}\left[\gbm{}{3} \cpoc -  \gbm{}{1}\dpod \right] \dot{\phi}\dot{\chi}\,,  
\eeq 
and
\beq
\label{eq:KGphi} 
\left( 1 + \doverc \ga^2 \rhochi \right) \ddot{\phi} + 3H\dot{\phi} \left( 1 - \doverc \ga^2 \pchi \right) + U' =  
\frac{1}{2}\left[\gbm{}{2}\rhochi + 3 \ga^2 \pchi \right]\cpoc - \frac{1}{2} \left(\ga^2 - 1\right) \rho_\chi \dpod \,.
\eeq
In the latter equation, the effective energy density $\rhochi$ and pressure $\pchi$ of the $\chi$ field can be expressed in the terms of the fields and their derivative using (\ref{eq:rhochi}) and (\ref{eq:pchi}), respectively.  

\subsubsection{Perturbations} \label{Sec:Pert}

To find equations of motion for the scalar perturbations on a cosmological background, we write,
\beqa
\phi = \phi(t) + \delta\phi(t,x^i) , \quad \chi = \chi(t) + \delta\chi(t,x^i) \, ,
\eeqa
and similarly perturb the FRW metric such that $g_{\mu\nu} = g^{(FRW)}_{\mu\nu} + \delta g_{\mu \nu}$. The background part $g^{(FRW)}_{\mu\nu}$ is given by (\ref{frw}), and  we write the metric perturbation for scalars, $\delta g_{\mu \nu}$,  explicitly in the Newtonian gauge as, 
\beq
\delta g_{\mu \nu} = diag(-2\Phi, -2 a(t)^2 \Psi, -2 a(t)^2 \Psi, -2 a(t)^2 \Psi) \, ,
\eeq
and expand the general equations of motion in section \ref{Sec:Action} to linear order in $\delta\phi$, $\delta\chi$, $\Phi$ and $\Psi$. From this procedure we obtain the field equations of the form,
\begin{align}
2 \left(\partial_i \partial^i \Psi - 3 H \dot{\Psi}\right)  = & \ \delta\rho = X_1 \Psi + X_2 \delta\phi + X_3 \dot{\delta\phi} + X_4 \delta\chi + X_5 \dot{\delta\chi} \, \label{eq:PEEtt},
\\ \nonumber \\
2 \left(\dot{\Psi} + H \Psi \right) =& \ -\delta q  =  - Y_1 \delta\phi + -Y_2 \delta \chi \,\label{eq:PEEts} ,
\\ \nonumber \\
2 \left(\ddot{\Psi} + 4H\dot{\Psi} + 4\dot{H}\Psi + 6H^2 \Psi \right) = & \ \delta p  = Z_1 \Psi + Z_2 \delta\phi + Z_3 \dot{\delta\phi} + Z_4 \delta\chi +Z_5 \dot{\delta\chi}\label{eq:PEEss} \, ,
\end{align}
where the coefficients $X_n$, $Y_n$, $Z_n$ are defined in terms of background quantities in appendix \ref{App:XYZ}. Note that in this process we have implicitly set $\Phi = \Psi$ using the $(i,j)$ component of the perturbed field equations. \\

We also find perturbed equations of motion for $\delta\phi$ and $\delta\chi$ of the general forms,
\beqa
\al{1} \delta \ddot{\phi} + \al{2} \delta \ddot{\chi} + \al{3}\partial_i \partial^i \delta\phi + \al{4}\partial_i \partial^i \delta\chi + \al{5} \dot{\Psi} + \al{6} \delta \dot{\phi} + \al{7} \delta \dot{\chi} + \al{8} \Psi + \al{9} \delta \phi + \al{10} \delta \chi  = 0 \, ,  \label{eq:alphas} 
\\ \nonumber \\
\be{1} \delta \ddot{\phi} + \be{2} \delta \ddot{\chi} + \be{3}\partial_i \partial^i \delta\phi + \be{4}\partial_i \partial^i \delta\chi + \be{5} \dot{\Psi} + \be{6} \delta \dot{\phi} + \be{7} \delta \dot{\chi} + \be{8} \Psi + \be{9} \delta \phi + \be{10} \delta \chi   = 0 \, , \label{eq:betas} 
\eeqa
where the coefficients $\al{n}$ and $\be{n}$ are explicitly written in appendix \ref{App:AlphaBeta}. \\

The system of equations (\ref{eq:PEEtt}--\ref{eq:betas}) can then be rewritten in terms of a new gauge-invariant variable to eliminate the explicit dependence on the metric perturbation $\Psi$. We proceed in the usual way by defining the gauge-invariant Sasaki-Mukhanov variables for our two fields such that,
\beqa
Q_\phi = \delta \phi + \frac{\dot{\phi}}{H} \Psi \, , \label{eq:SMphi} \\
Q_\chi = \delta \chi + \frac{\dot{\chi}}{H} \Psi \, . \label{eq:SMchi}
\eeqa
Substituting these definitions into eqs. (\ref{eq:alphas}--\ref{eq:betas}) and using eqs. (\ref{eq:PEEtt}--\ref{eq:PEEss}) to eliminate derivatives of $\Psi$ we can rewrite the system in the form,
\beqa
\al{1} \ddot{Q}_\phi + \al{2} \ddot{Q}_\chi +  \al{3}\partial_i \partial^i Q_\phi  +  \al{4}\partial_i \partial^i Q_\chi  + \alb{6} \dot{Q}_\phi  + \alb{7} \dot{Q}_\chi  + \alb{9} Q_\phi + \alb{10} Q_\chi  = 0 \, ,  \label{eq:Qal} 
\\ \nonumber \\
\be{1} \ddot{Q}_\phi  + \be{2} \ddot{Q}_\chi + \be{3}\partial_i \partial^i Q_\phi + \be{4}\partial_i \partial^i Q_\chi + \beb{6} \dot{Q}_\phi  + \beb{7} \dot{Q}_\chi + \beb{9} Q_\phi + \beb{10} Q_\chi   = 0 \, , \label{eq:Qbe} 
\eeqa
where the new $\alb{n}$ and $\beb{n}$ coefficients are given in appendix \ref{App:AlphaBetaBar}. Note that for $n<5$ the $\al{n}$ and $\be{n}$ coefficients are unchanged as the perturbed field equations for $\Psi$ contain no second derivatives of the scalar fields. It is this system of equations (\ref{eq:Qal}--\ref{eq:Qbe}) that we will proceed to solve in section \ref{Sec:ResultsBG} to obtain inflationary predictions. \\

As we have chosen to write our theory such that the gravity sector is unmodified, the equations governing the dynamics of the tensor perturbations are identical to the ordinary case.

\section{The two-field system} \label{Sec:Numerics}

In this Section we will analyse the equations of motion of our coupled two-field system. In particular, we will extract the canonical, gauge-invariant degrees of freedom in the scalar fluctuations about the cosmological background, deduce the physical properties of the propagating fields from the resulting equations of motion for the suitable variables, and find the natural initial conditions for inflation to set into these equations. Due to the excessively nonlinear nature of the background equations (\ref{eq:EE1}--\ref{eq:KGphi}), and the sheer complexity of the perturbation equations (\ref{eq:Qal}--\ref{eq:Qbe}), analytical methods are of little help in understanding the full physics of our system. We hence resorted to a full numerical study of the two-field system  \cite{Tsujikawa:2003ccc} in the following Section \ref{Sec:ResultsBG}. The stage is set for the numerical methodology in this section. \\

To calculate the power spectra, we Fourier transform the equations, such that,
\beqa
\al{1} \ddot{Q}_\phi + \al{2} \ddot{Q}_\chi -  \al{3} \frac{k^2}{a^2} Q_\phi  -  \al{4}\frac{k^2}{a^2} Q_\chi  + \alb{6} \dot{Q}_\phi  + \alb{7} \dot{Q}_\chi  + \alb{9} Q_\phi + \alb{10} Q_\chi  = 0 \, ,  \label{eq:kQal} 
\\ \nonumber \\
\be{1} \ddot{Q}_\phi  + \be{2} \ddot{Q}_\chi - \be{3}\frac{k^2}{a^2} Q_\phi - \be{4}\frac{k^2}{a^2} Q_\chi + \beb{6} \dot{Q}_\phi  + \beb{7} \dot{Q}_\chi + \beb{9} Q_\phi + \beb{10} Q_\chi   = 0 \, , \label{eq:kQbe} 
\eeqa
and then solve these equations for a range of $k$ values. For best compatibility with observational data, we follow the method of the latest Planck paper to determine the location of the observable window during inflation, and hence find the range of $k$ values which we need to integrate the system for to calculate observables. We take the range of observable scales to be $k \in [10^{-3},10^{1}]$ \si{\per\mega\parsec}.  

\subsection{Identification of sound speeds} \label{Sec:Sound}

For uncoupled two-field systems, one can identify the sound speeds of the perturbations from the ratio of the prefactors of the second spatial and temporal derivatives, and for canonic fields this will be identically equal to the speed of light, $c_s = 1$, in Planck units. However, with two interacting fields, the perturbed equations generally contain mixtures of the second derivatives of both perturbed fields, and so it is important to identify the linear combinations of the two fields which have a single propagation speed --- the fundamental degrees of freedom. We find that, because $\al{2}$ and $\al{4}$ in our model are equal to zero, $Q_\phi$ is a fundamental degree of freedom which propagates with a sound speed,
\beq \label{eq:csphi}
c^{(\phi)}_s = \sqrt{-\frac{\al{3}}{\al{1}}} = \sqrt{\frac{C - D \ga^2 \pchi}{C + D \ga^2 \rhochi}} \, .
\eeq
However, as $\be{1}$ and $\be{3}$ are non-zero in general, we find that $Q_\chi$ is not fundamental and that instead it is the variable
\beq \label{eq:qtheta}
Q_\theta = \be{2} Q_\chi + \be{1} Q_\phi = Q_\chi + \ga^2 \doverc \phidot \chidot Q_\phi \, ,
\eeq
which propagates with a sound speed of
\beq \label{eq:cstheta}
c^{(\theta)}_s = \sqrt{-\frac{\be{4}}{\be{2}}} = \frac{1}{\ga} \, .
\eeq
We can then rewrite the system of equations (eq. \ref{eq:Qal}--\ref{eq:Qbe}) in matrix form as,
\beq
\ddot{\mathcal{Q}} + C^2_s \frac{k^2}{a^2} \mathcal{Q} + \pmat{ \left(\alb{6} - \be{1}\alb{7}\right)/\al{1}  &  \alb{7}/\al{1} \\ \beb{6} - \be{1}\beb{7} - 2 \dot{\be{1}} & \beb{7}  } \dot{\mathcal{Q}}+ \pmat{ \left(\alb{9} - \be{1}\alb{10} - \dot{\be{1}}\alb{7} \right)/\al{1}  &  \alb{10}/\al{1} \\ \beb{9} - \be{1}\beb{10} - \dot{\be{1}}\beb{7} - \ddot{\be{1}} & \beb{10}  } \mathcal{Q} = 0 \, , \label{eq:MatrixPert}
\eeq
where
\beq
\mathcal{Q} = \pmat{Q_\phi \\ Q_\theta}, \quad C_s =  \pmat{c^{(\phi)}_s & 0 \\ 0  & c^{(\theta)}_s} ,
\eeq
and we have used the fact that $\al{2}=\al{4}=0$ and $\be{2}=1$ in our model to simplify the matrix coefficients. \\

We find one scalar mode with the sound speed $c^{(\theta)}_s=1/\gamma$, as expected, since the propagation of the field $\chi$ is slowed down by the disformal coupling; in the brane world picture of Section \ref{Sec:String} this would be interpreted as the Lorenz dilation due to the brane movement, $c_s \rightarrow c_s/\gamma$. The same phenomenon was noted to occur also for a vector field residing upon a brane \cite{Koivisto:2014gia,Koivisto:2015vda}. The sound speed of the other degree of freedom, $c^{(\phi)}_s$, turns out to assume the more nontrivial form (\ref{eq:csphi}). Note that the physical propagation speed is a frame-independent quantity \cite{Domenech:2015hka}.

\subsection{Initial conditions for perturbations} \label{Sec:ICs}

To find appropriate initial conditions for numerical integration, we need to find a solution of eq. (\ref{eq:MatrixPert}) in the early time limit. In this limit, we assume that the coefficients $\al{n}$, $\be{n}$ are constant and the expansion of space is de Sitter-like so $H$ can also be assumed constant. Finally, we assume that interactions can be neglected, which is a good approximation when the off-diagonal components of the matrices in eq. \ref{eq:MatrixPert} are smaller than the diagonal components. We will later, when we look at example trajectories, discuss whether this condition for the validity of the approximation is well-fulfilled in each case. Working in conformal time, $\bd \eta = a(\eta) \bd t$ and using the canonically normalised variables $\upsilon_\phi = a(\eta) Q_\phi$ and $\upsilon_\theta = a(\eta) Q_\theta$, the equations we are trying to solve are,
\beqa
\eta^2 \al{1} \upsilon_{\phi}'' + \eta \left(3 \al{1} - \frac{\alb{6} - \be{1}\alb{7}}{H}\right) \upsilon_{\phi}' + \left(- \al{3}  k^2 \eta^2 + \al{1} + \frac{ \left(\alb{9} - \be{1}\alb{10} - \dot{\be{1}}\alb{7} \right) - H  \left(\alb{6} - \be{1}\alb{7}\right)}{H^2}\right) \upsilon_{\phi} = 0 \, , \nonumber \\ 
\eta^2 \be{2} \upsilon_{\theta}'' + \eta \left(3 \be{2} - \frac{\beb{7}}{H}\right) \upsilon_{\theta}' + \left(- \be{4}  k^2 \eta^2 + \be{2} + \frac{\beb{10} - H \beb{7}}{H^2}\right) \upsilon_{\theta} = 0 \, . \nonumber
\eeqa
These equations can be solved in terms of Hankel functions with the expressions,
\beqa
\upsilon_{\phi} = \left(-\eta\right)^{\frac{1}{2}+ \xi_\phi} \left[ C_1 H^{(1)}_{\Omega_\phi}(- c^{(\phi)}_s k \eta) + C_2 H^{(2)}_{\Omega_\phi}(- c^{(\phi)}_s k \eta) \right]\,, \label{eq:Hankelphi} \\
\upsilon_{\theta} = \left(-\eta\right)^{\frac{1}{2}+ \xi_\theta} \left[ C_3 H^{(1)}_{\Omega_\theta}(- c^{(\theta)}_s k \eta) + C_4 H^{(2)}_{\Omega_\theta}(- c^{(\theta)}_s k \eta) \right]\,, \label{eq:Hankeltheta}
\eeqa
where
\begin{align*}
\xi_\phi & = \ \frac{\alb{6} - \be{1}\alb{7}}{2 H \al{1}} - \frac{3}{2} \, , \\
\xi_\theta & = \  \frac{\beb{7}}{2 H \be{2}} - \frac{3}{2} \, , \\
\Omega_\phi & = \  \left(\frac{\alb{6} - \be{1}\alb{7}}{2 H \al{1}}\right)^2 - \left(\frac{\alb{9} - \be{1}\alb{10} - \dot{\be{1}}\alb{7}}{H \al{1}}\right)^2 \, , \\
\Omega_\theta & = \ \left(\frac{\beb{7}}{2 H \be{2}}\right)^2 - \left(\frac{\beb{10}}{H \be{2}}\right)^2  \, .
\end{align*}
Conventionally, one would now take the asymptotic behaviour of these solutions as $\eta \rightarrow -\infty$ and choose the constants of integration $C_i$ to make the solutions match the Bunch-Davies vacuum. However, this procedure only works in the standard case because the parameters $\xi_\phi$ and $\xi_\theta$ are zero for ordinary inflation (for which $\alb{6} = \beb{7} = 3H$ and $\al{1} = \be{2} = 1$, in particular). The solutions of our model, eqs. (\ref{eq:Hankelphi}--\ref{eq:Hankeltheta}), contain an extra factor $(-\eta)^{\xi_I}$ ($I = \phi,\theta$)  in comparison with the uncoupled ($D=0$) case, and this remains when taking the early time asymptotic behaviour of the solution. The parameters $\Omega_I$ also generally differ from the uncoupled case but this deviation can be absorbed in our choice of integration constants. In summary, we can choose initial conditions which are as close to the Bunch-Davies state as possible, but contain an additional time-dependent factor, that is, 
\beqa
Q_\phi \left(\eta \rightarrow -\infty\right) =  \frac{e^{-i c^{(\phi)}_s k \eta}}{\sqrt{2 c^{(\phi)}_s k} a} \left(-\eta\right)^{\frac{\alb{6} - \be{1}\alb{7}}{2 H \al{1}}- \frac{3}{2}} \, , \label{eq:ICphi}
\\
Q_\theta \left(\eta \rightarrow -\infty\right) =  \frac{e^{-i c^{(\theta)}_s k \eta}}{\sqrt{2 c^{(\theta)}_s k} a} \left(-\eta\right)^{\frac{\beb{7}}{2 H}- \frac{3}{2}} \, . \label{eq:ICtheta}
\eeqa
Finally, as we will still solve the system eqs. (\ref{eq:Qal}--\ref{eq:Qbe}) in terms of $Q_\chi = Q_\theta - \ga^2 \doverc Q_\phi$, we also identify the initial condition for this as a combination of the above two initial conditions, 
\beq
Q_\chi \left(\eta \rightarrow -\infty\right) = \frac{e^{-i c^{(\theta)}_s k \eta}}{\sqrt{2 c^{(\theta)}_s k} a} \left(-\eta\right)^{\frac{\beb{7}}{2 H}- \frac{3}{2}} - \frac{ \ga^2 \doverc e^{-i c^{(\phi)}_s k \eta}}{\sqrt{2 c^{(\phi)}_s k} a} \left(-\eta\right)^{\frac{\alb{6} - \be{1}\alb{7}}{2 H \al{1}}- \frac{3}{2}} \, . \label{eq:ICchi}
\eeq
It is also useful to begin integration of the differential equations at a time later than $N = 0$, as at this point, the perturbations are rapidly oscillating functions which cannot be properly resolved without using excessively small time-steps. At the same time, however, it is important to not begin integration too late, else the assumptions in finding valid initial conditions become poor, and accuracy will be reduced. We define the parameter $x$, such that,
\beq
c^{(x)}_s k = x a H \, ,
\eeq
where $c^{(x)}_s$ is the smaller of the two sound speeds at a given time. When $x=1$, the fields are in the middle of horizon crossing, and the oscillatory initial conditions are no longer appropriate. It is therefore important to begin integration at a time when $x \gg 1$. We typically begin integration at the time when $x = 50$. \\

\subsection{Calculating the power spectrum}

In order to calculate the power spectrum of the adiabatic perturbations, we must first obtain the comoving curvature perturbation $\mathcal{R}$. This is given by the relation, \cite{Gordon:2000hv}

\beq
\mathcal{R} = \Psi - \frac{H}{\rho_\phi + \rho_\chi + p_\phi + p_\chi} \delta q \, ,
\eeq

The use of this standard result is justified because we are analysing our system entirely in terms of the gravitational metric $g$, such that the gravity sector of our theory is unmodified GR. Using the form of the momentum perturbation $\delta q$ obtained in eq. \ref{eq:PEEts} and the definition of the Sasaki-Mukhanov variables (eqs. \ref{eq:SMphi}--\ref{eq:SMchi}), this can be recast in the gauge-invariant form, 

\beq
\mathcal{R} = - \frac{H}{\rho_\phi + \rho_\chi + p_\phi + p_\chi} \left[ Y_1 Q_\phi + Y_2 Q_\chi \right] =  \frac{H}{\rho_\phi + \rho_\chi + p_\phi + p_\chi} \left[ \left(1 + \frac{D}{C} \rho_\chi \right) \phidot Q_\phi + \gamma C \chidot Q_\chi \right] \, .
\eeq


We can further rewrite this expression to be in terms of the canonical variable $Q_\theta$ identified in section \ref{Sec:Sound}, using eq. \ref{eq:qtheta}, so that,

\beq
\mathcal{R} =  \frac{H}{\rho_\phi + \rho_\chi + p_\phi + p_\chi} \left[ \left(1 + \frac{D}{C} \rho_\chi - \gamma^3 D \chidot^2 \right) \phidot Q_\phi + \gamma C \chidot Q_\theta \right] \, .
\eeq

This is a generalisation of the usual result for two typical scalar fields, with additional disformal coupling terms. From this we can then obtain the power spectrum with the usual definition, $2 \pi^2 \mathcal{P}_\mathcal{R} = k^3 | \mathcal{R} |^2$. We could similarly proceed to calculate the entropy perturbations, with the analogous relation, 

\beq
\mathcal{S} = H \left( \frac{\delta p}{\dot{p_\phi} + \dot{p_\chi}} - \frac{\delta \rho}{\dot{\rho_\phi} + \dot{\rho_\chi} } \right) \, ,
\eeq

but elect to leave a comprehensive study of the non-adiabatic spectrum of disformally coupled inflation to a future work, and focus in this work on exploring the standard adiabatic spectrum and its properties. 

\section{Inflationary predictions} \label{Sec:ResultsBG}

We now move on to study the dynamics of the system. For simplicity and in order to add as few additional parameters to the model as possible, we choose standard quadratic massive field potentials, that is,

\beq
U(\phi) = \frac{1}{2} m_\phi^2 \phi^2 \, , \quad V(\chi) = \frac{1}{2} m_\chi^2 \chi^2 \, ,
\eeq

and numerically investigate the consequences of the disformal coupling in the content of inflation. We note that as $\ga$ (recall eq. (\ref{eq:gamma}) for the definition) is trivially equal to $\ga=1$ when disformal coupling is absent, and that as it increases the disformal terms in the equations of motion become more significant, it is a useful parameter to characterise the importance of disformal effects at a given time. \\

We will first consider the power-law forms for the couplings in part \ref{Sec:BG_String}, where we do not find inflation emerging naturally without fine-tuning. In part \ref{Sec:BG_Phenom} we employ exponential couplings and enumerate 4 types of dynamical evolutions that could describe viable inflation. 

\subsection{System with power-law couplings} \label{Sec:BG_String}

With coupling functions given by the braneworld scenario of section \ref{Sec:String}, eqs. (\ref{eq:StringC}--\ref{eq:StringD}), we find for a wide range of parameters and initial conditions it is not possible to produce inflation in the presence of non-trivial disformal effects. That is, if inflation occurs, $\ga$ is very close to $1$ and there is little to no effect from the disformal parts of the equations of motion. Similarly, if disformal effects are large enough, then inflation does not happen at all. For some sets of parameters, we observed backgrounds which started off with large $\ga$  but no inflation, and then settled into a period of inflation, but only after $\ga$ had decreased to be very close to $1$ again. Additionally, some trajectories where $\ga$ becomes large only as inflation ends (far past the observable window) were observed, and may have interesting effects on reheating that are beyond the scope of this present work.\\

The reasons why inflation fails to occur in the presence of disformal effects for this choice of coupling are split into two categories; those which prevent $\phi$ from serving as an inflation, and those which similarly inhibit $\chi$'s action as an inflaton. Note that while these observations do not totally forbid simultaneous inflationary and disformal behaviour, numerical checks have been carried out on various exceptional sets of parameters, verifying that they also do not produce interesting modifications to inflation.

\subsubsection{$\phi$ as the inflaton} \label{Sec:StringPhi}

As we have $C = T_3^{-\frac{1}{2}} \phi^2$ and $D = T_3^{-\frac{1}{2}} \phi^{-2}$, the equation for $\ga$ can be written as,
\beq
\label{eq:StringGamma}
\ga = \left(1 - \frac{\phidot^2}{\phi^4}\right)^{-\frac{1}{2}} \, .
\eeq
However, $\phi$ is a standard minimally coupled field in this model. If it is to function as an inflaton then it must satisfy the slow-roll condition $\phidot^2 \ll U$. For a massive field potential, $U \propto m_\phi^2\phi^2$, so we have that $\phidot^2 \ll m_\phi^2\phi^2$. Applying this condition to equation (\ref{eq:StringGamma}) indicates that $\ga$ will be strongly suppressed while $\phi$ is on an inflationary trajectory. We can therefore see that the slow-roll condition for inflation to occur with $\phi$ as the inflaton is incompatible with the conditions for non-trivial disformal behaviour. \\

This observation is related to why analytical approaches to studying this system were of little use --- the slow-roll approximation is a key tool in the analysis of inflationary models but it cannot account for disformal effects as it automatically neglects them. Inspection of the equations of motion reveals that the disformal terms often contain factors of $\phidot$, $\chidot$ and $\gbm{}{1}$ which are all negligible in a slow-roll approximation. We were not able to find an alternative approximation scheme capable of producing analytical results in good agreement with the full numerical solutions. \\

\subsubsection{$\chi$ as the inflaton}  \label{Sec:StringChi}

It would alternatively be possible to set up $\chi$ to be an inflaton, and have $\phi$ remain small to allow $\ga$ to take on large values. However this encounters problems in the form of the conformal coupling in the effective energy density of $\chi$ eq. (\ref{eq:rhochi}),
\beq 
\rhochi = \gamma T_3^{-\frac{1}{2}} \phi^2 \left( \frac{1}{2}\gamma^2\dot{\chi}^2+ T_3^{-\frac{1}{2}} \phi^2 V\right)  \, ,
\eeq
which, for small $\phi$, is generally suppressed, inhibiting $\chi$'s function as an inflaton. 

Another concern is generic in models where $\chi$ an inflaton and, regardless of the form of the couplings, arises from that if we take the expressions for the energy density and pressure of $\chi$ (eqs. \ref{eq:rhochi}--\ref{eq:pchi}), we find its equation of state to be,

\beq \label{eq:wchi}
w_\chi = \frac{\pchi}{\rhochi} = \ga^{-2} \frac{\ga^2 \chidot^2 - 2 C V}{\ga^2 \chidot^2 + 2 C V} \, .
\eeq

However, the condition for a field to function as an inflaton is $w < -1/3$. This situation is inhibited by the factor of $\ga^{-2}$ in equation (\ref{eq:wchi}), which, in the presence of large $\ga$, drives the equation of state to $0$, and leading to non-inflationary expansion. Furthermore, a scalar field with a strongly negative equation of state must be dominated by its potential energy, and the kinetic energy terms in eq. (\ref{eq:wchi}) are proportional to $\ga^2$. Even assuming perfect potential energy domination, $w_\chi = -\ga^{-2}$, and hence $\ga$ must be less than $\sqrt{3}$ for $\chi$ to drive inflation. While $\ga = \sqrt{3}$ would be far enough from $\ga = 1$ to affect the dynamics, we find numerically again that such a state is not stable and the dynamics quickly transition into a period of inflation with trivial $\ga$, or that little to no inflation occurs in such scenarios.

\subsection{System with exponential couplings} \label{Sec:BG_Phenom}

The phenomenological coupling functions introduced in section \ref{Sec:Phenom}, eqs. (\ref{eq:PhenomC}--\ref{eq:PhenomD}), are, however, much more flexible and allow us to find models where inflationary trajectories and significant disformal effects are compatible. With these couplings, the explicit form of $\ga$ is now,

\beq
\label{eq:PhenomGamma}
\ga = \left(1 - \frac{D_0}{C_0}e^{(d-c)\phi}\phidot^2\right)^{-\frac{1}{2}} \, ,
\eeq

which, for a sufficiently large value of $D_0/C_0$ and a sufficiently positive value of $(d-c)$, is capable of sustaining a non-trivial value on an inflationary trajectory. Such a choice of parameters enables us to avoid the problems discussed in section \ref{Sec:StringPhi} and hence allows $\phi$ to function as an inflaton even in the presence of disformal effects. In the following sections, we present a set of interesting behaviours exhibited for exponentially coupled fields. In all of these examples, $\phi$ is the dominant source of energy density. A generic result obtained from trajectories with non-trivial disformal effects is the boosting of the scalar amplitude due to subluminal sound speeds during the observable window \cite{Nakashima:2011erf}, and hence a corresponding lowering of the tensor-to-scalar ratio. Albeit less generally, it was also found that disformal effects could produce trajectories with sharp turns, which we believe could in certain cases produce features in the power spectrum \cite{Palma:2015eth,Mizuno:2014sfg}, and hence potentially construct observational tests for disformal couplings, though this possibility was not focused on in this exploratory work. Further features could be induced by disformal effects via variations in the sound speeds \cite{Achucarro:2013jte,Achucarro:2014tj}. Furthermore, in the following examples we shall tend to restrict ourselves to simplistic choices of the masses of the fields, typically choosing them to be equal. While allowing a more general mass hierarchy would likely increase the range of interesting evolutions obtainable, in the interest of concentrating on effects of disformal origins we shall refrain from making use of this. \\

Here we also choose to restrict $D_0$ to positive values. As discussed in section \ref{Sec:Background}, negative choices of $D$ corresponds to $\ga < 1$. While this choice does not exclude general inflationary behaviour, we find several undesirable properties in this regime. Firstly, from the expressions for the sound-speeds of the two scalar perturbations derived in section \ref{Sec:Sound} it is clear that $\ga < 1$ leads to superluminal propagation speeds. Similarly, from eq. (\ref{eq:wchi}) we can see that this would lead to $w_\chi < -1$. These non-standard conditions are physically unappealing, but we also observe that in contrast to $D > 0$ where the modified scalar propagation speed produces a useful reduction of the tensor-to-scalar-ratio $r$, the $D < 0$ case instead enhances it. As present observations require a small value of $r$, this effect is phenomenologically undesirable, particularly in combination with the aforementioned non-standard behaviours of the two $c_s$s and $w_\chi$. \\

For the trajectories shown here, we find that the choices of parameters and initial field values are compatible with the assumptions made in our scheme for setting initial conditions for the perturbations. In particular, we focus on the elements of the matrix which multiplies $\dot{\mathcal{Q}}$ in eq. \ref{eq:MatrixPert}, as the values of these functions are found numerically (verified for trajectories A--D) to be typically around six order of magnitude larger than those in the effective mass matrix multiplying $\mathcal{Q}$ at early times. Additionally, the diagonal elements of this matrix are found to be $\mathcal{O}(10^{2})$ (trajectory C) or $\mathcal{O}(10^{4})$ (trajectories A,B and D) times larger than the off-diagonal elements at early times, justifying the assumptions made in section \ref{Sec:Sound} regarding decoupling of the two fields at early times and hence our approximate initial conditions. 


\subsubsection{Trajectory A} \label{Sec:BG_A}

\begin{figure}[t]
    \centering
    \includegraphics[width=\textwidth]{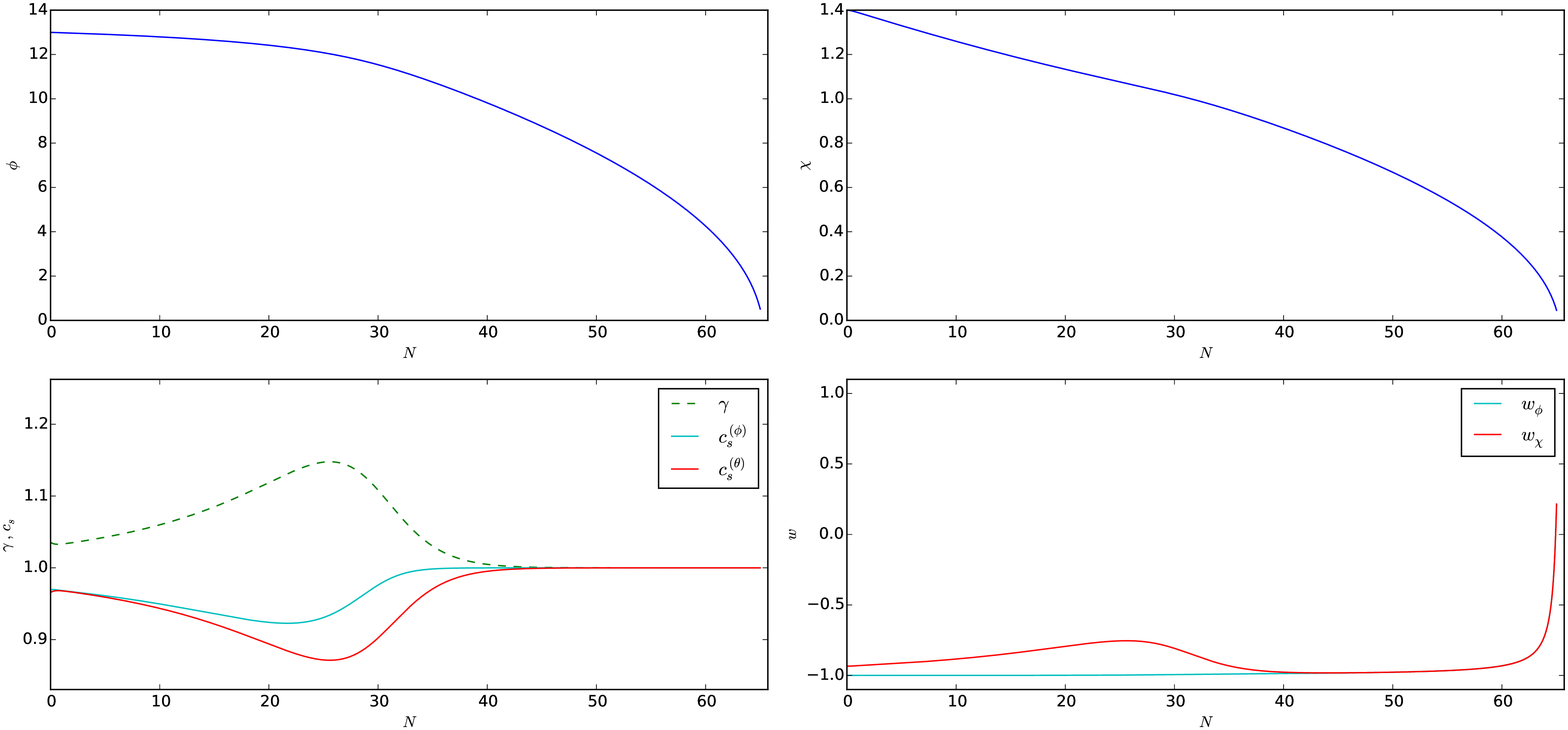}
     \includegraphics[width=\textwidth]{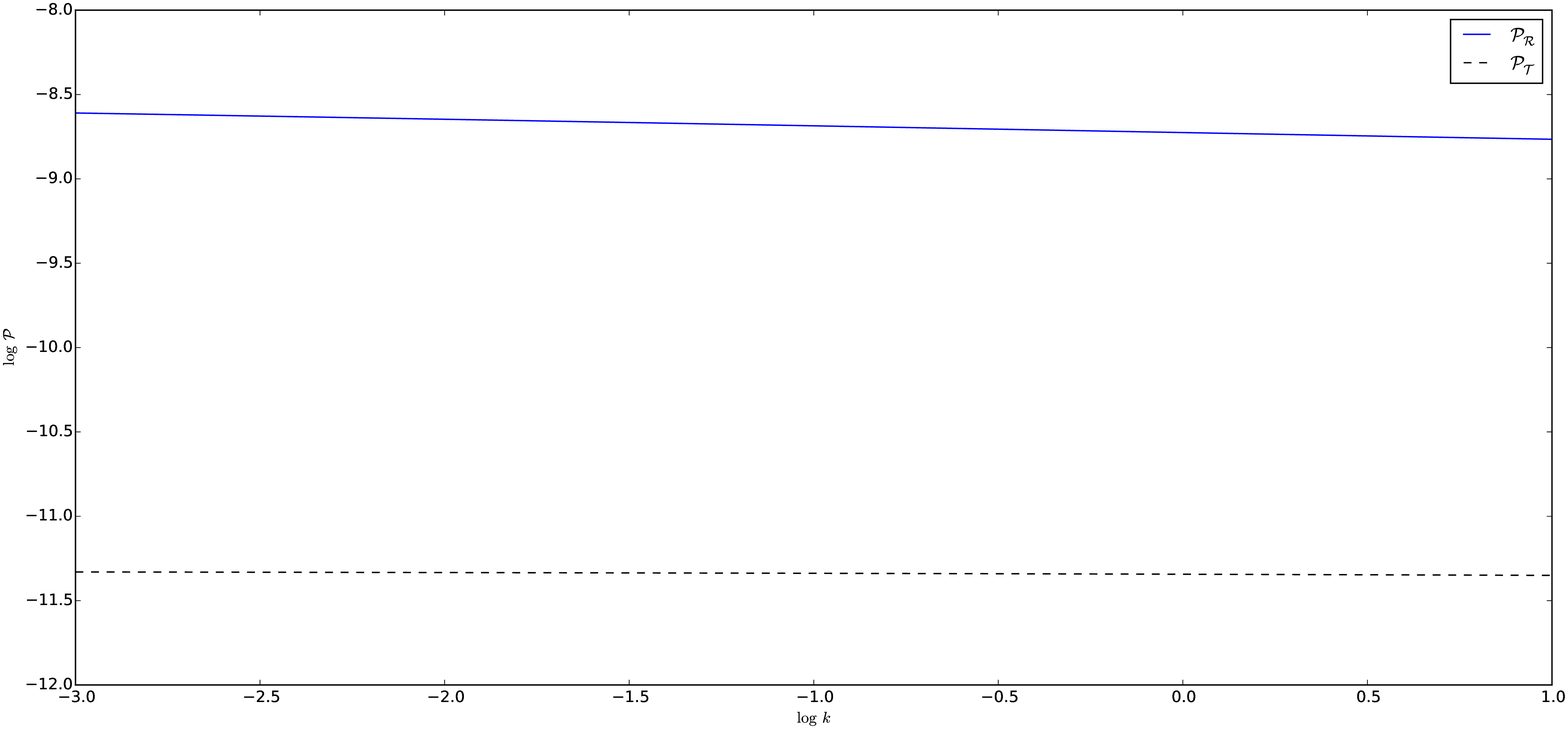}
    \caption{Trajectory A's background evolution of fields, sound speeds and equations of state as a function of the e-fold number $N$ are shown in the upper four graphs. Parameters and initial conditions used are: $d = 2$, $c = 0$, $D_0 = 13.5$, $C_0 = 1$, $m_\phi = m_\chi = 1.8 \times 10^{-6}$, $\phi_0 = 13$, $\chi_0 = 1.4$, $\phidot_0 = -1.7 \times 10^{-7}$, $\chidot_0 = 0$; see section \ref{Sec:BG_A}. The lower plot shows the scalar ($\mathcal{P}_\mathcal{R}$, blue solid line) and tensor ($\mathcal{P}_\mathcal{T}$, black dashed line) power spectra for this trajectory.}
    \label{fig:BGA}
\end{figure}

Shown in figure \ref{fig:BGA}, the evolution with this set of parameters has two distinct parts. First, a period of around 25 e-folds where $\ga$ is increasing up to a maximum, before a second period where $\ga$ decreases to $1$ and remains there until the end of inflation. This kind of behaviour is found when a large positive value of $(d-c)$ is the source of $\ga$'s departure from unity.  At first, $\phi$ begins to roll down its potential, causing $\phidot$ to increase quickly enough to lead to an overall increase in $\ga$, but as $\phi$ decreases, the steep exponential factor in $\ga$ also decreases as the value of $\phidot$ begins to stabilise, leading to an eventual decrease in $\ga$. This exponential levelling-off of $\ga$ at late times also suggests that the disformal coupling will have become trivial by the time reheating begins, allowing it to proceed in the standard way.

The parameters for trajectory A produce a power spectrum with $A_s = 2.12 \times 10^{-9}$ and $n_s = 0.961$ at the Planck pivot scale, $k_{pivot} = 0.05$ \si{\per\mega\parsec} and a tensor-to-scalar ratio of $r_{0.002} = 0.0172$, as shown in the lower graph in figure \ref{fig:BGA} (for the definitions of the observables, see  \cite{Ade:2015lrj}). The scalar perturbations are amplified with respect to ordinary models of inflation, so we have a much smaller value of $r$ than the standard case. Running of the spectral index is calculated to be $\alpha=-4 \times 10^{-4}$. 



\subsubsection{Trajectory B} \label{Sec:BG_B}

In figure \ref{fig:BGB} we show a trajectory where the large value of $D_0/C_0$ is the primary source of an increased $\ga$. In this kind of situation, $\ga$ typically starts at some maximum value and then smoothly decreases as inflation progresses, corresponding to the exponential factor decreasing as $\phi$ decreases, but as $\phidot$ remains fairly constant. In this kind of model, we also observe that $\chi$ is near-constant at early times, and that this is due to the effects of the conformal coupling --- a more negative value of $c$ is numerically observed to hold $\chi$ constant more effectively, and increase in turn the sharpness of the trajectory in $\phi$-$\chi$ space. Once again we find that this type of model reduces to trivial disformal coupling with $\ga=1$ at late times.

\begin{figure}
    \centering
    \includegraphics[width=\textwidth]{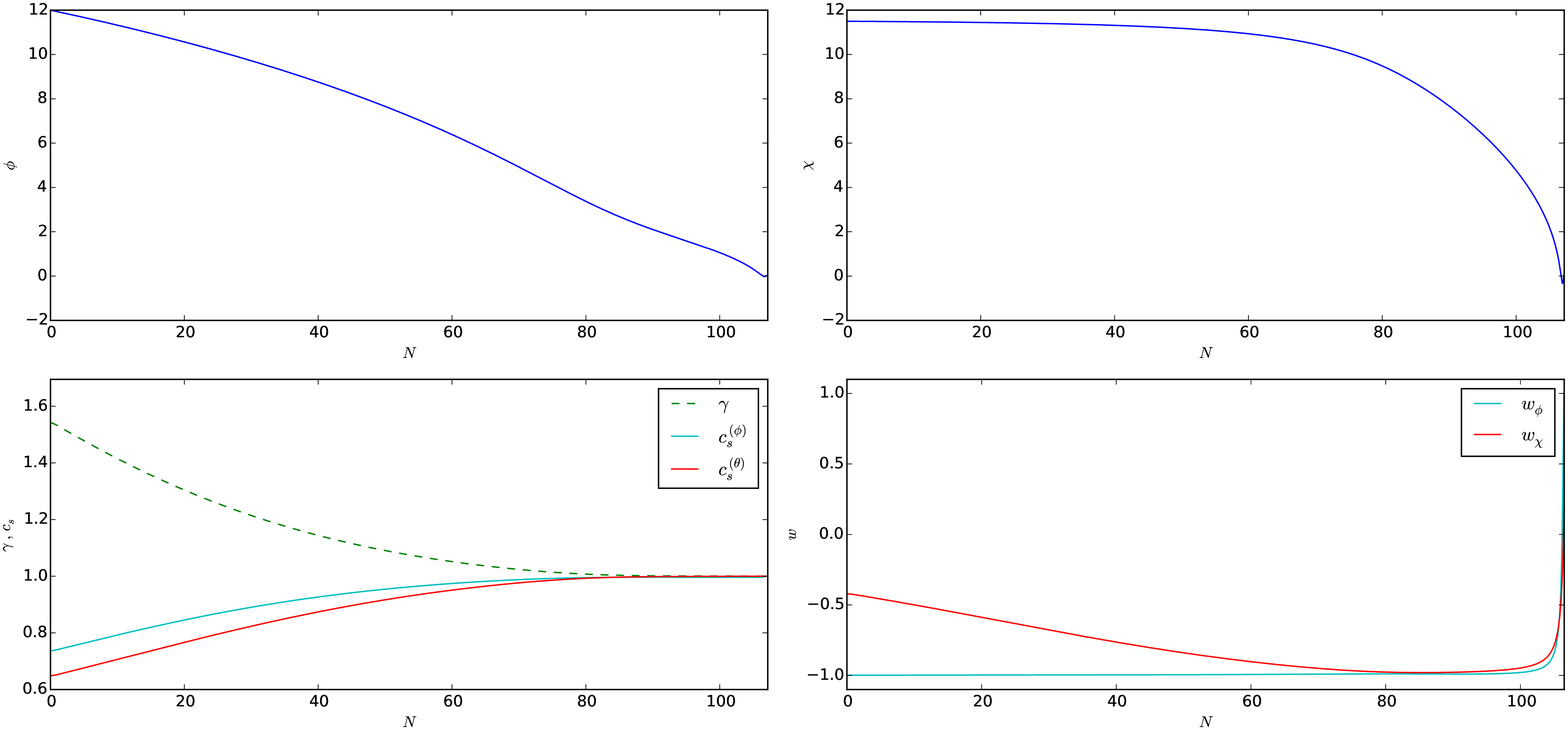}
    \includegraphics[width=\textwidth]{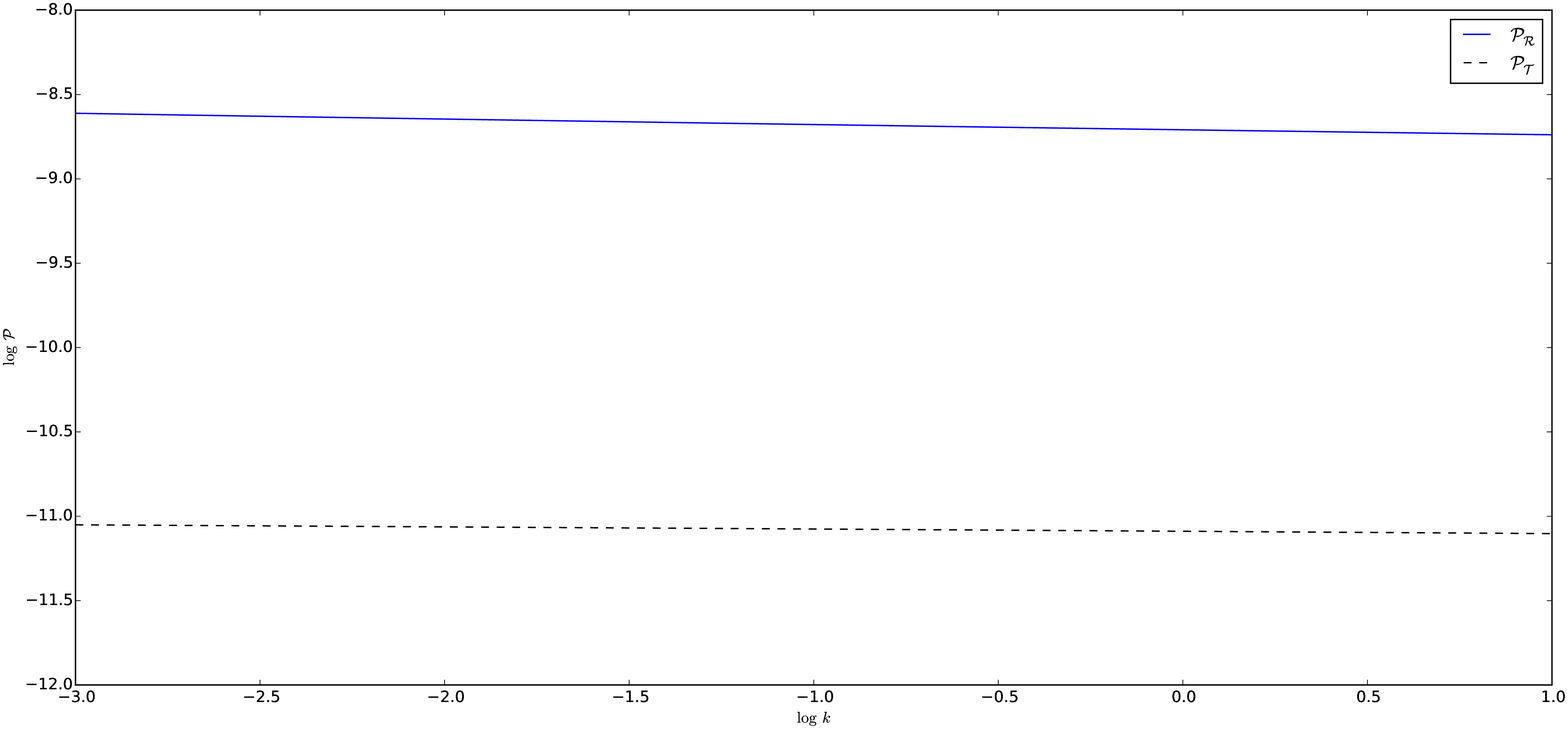}
    \caption{Trajectory B's background evolution of fields, sound speeds and equations of state as a function of the e-fold number $N$ are shown in the upper four graphs.  Parameters and initial conditions used are: $d = 0.19$, $c = -0.19$, $D_0 = 8 \times 10^9$, $C_0 = 1$, $m_\chi = 1.39 \times 10^{-6}$, $m_\phi = 2.78 \times 10^{-6}$ , $\phi_0 = 12$, $\chi_0 = 11.5$, $\phidot_0 = -8.7 \times 10^{-7}$, $\chidot_0 = 0$; see section \ref{Sec:BG_B}. The lower plot shows the scalar ($\mathcal{P}_\mathcal{R}$, blue solid line) and tensor ($\mathcal{P}_\mathcal{T}$, black dashed line) power spectra for this trajectory.}
    \label{fig:BGB}
\end{figure}

Similarly to trajectory A, the boosting of the scalar spectrum in due to disformal effects means that a spectrum with properties consistent with experiment,  $A_s = 2.15 \times 10^{-9}$ and $n_s = 0.968$, has a tensor-to-scalar ratio of just $r_{0.002} = 0.0314$. This is shown in the lower plot of figure \ref{fig:BGB}. We find that the running of the spectral index is $\alpha=7 \times 10^{-4}$. 



\subsubsection{Trajectory C} \label{Sec:BG_C}

Usually, scalar fields must have super-Planckian initial values in order to drive a useful amount of inflation. In this model, shown in figure \ref{fig:BGC}, we instead find that fields with sub-Planckian initial conditions can drive large amounts of inflation in the presence of a large enough disformal coupling. In this example we use $D = 5.00 \times 10^{21}$, which corresponds to a mass scale of $m_D = 3.76 \times 10^{-6}$. Here, $\ga$ takes on an extremely large value, and remains fairly constant until the end of inflation where it sharply decreases. With this trajectory, however, we observe that after inflation, $\ga$ undergoes oscillations corresponding to the post-inflationary oscillations of $\phi$ around its minimum, as now the disformal coupling is so large that even these small oscillations produce a non-trivial change in $\ga$ - this hence may lead to effects during reheating that are beyond the scope of this present work. Due to the size of $\ga$, both sound speeds are close to zero for this trajectory.

\begin{figure}
    \centering
    \includegraphics[width=\textwidth]{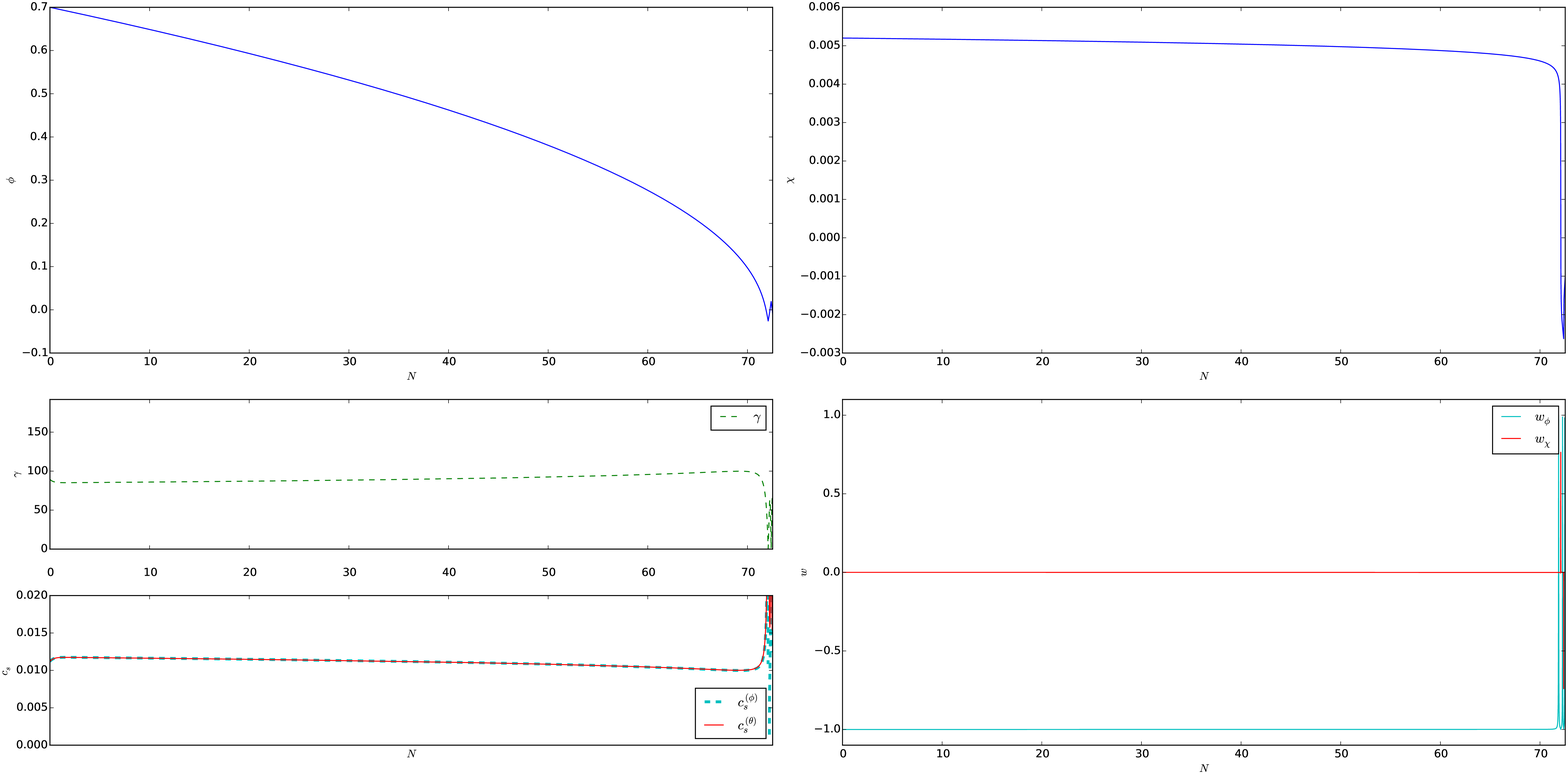}
    \includegraphics[width=\textwidth]{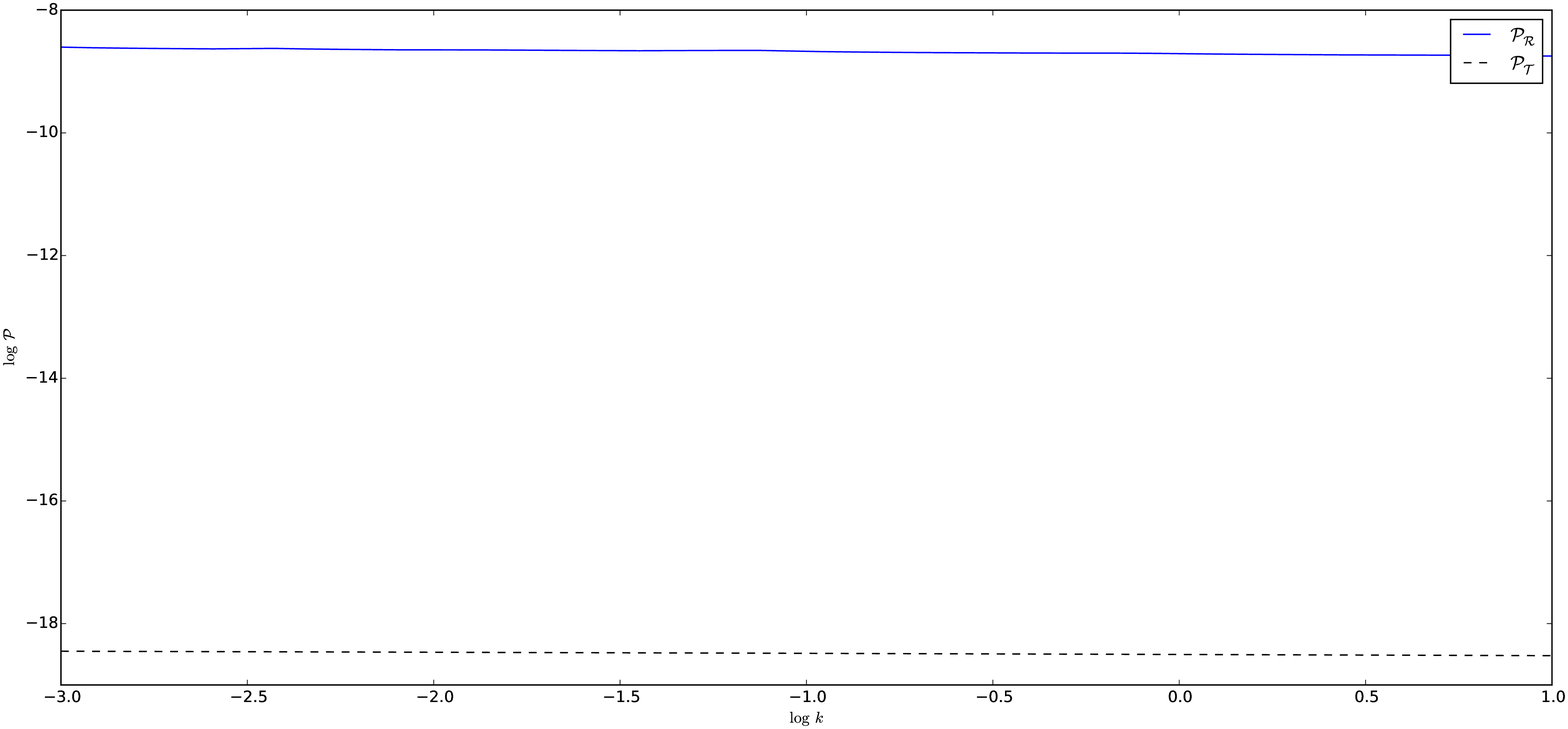}
    \caption{Trajectory C's background evolution of fields, sound speeds and equations of state as a function of the e-fold number $N$ are shown in the four upper graphs. Parameters and initial conditions used are: $d = 0$, $c = 0$, $D_0 = 5 \times 10^{21}$, $C_0 = 1$, $m_\phi = m_\chi = 1 \times 10^{-8}$, $\phi_0 = 0.7$, $\chi_0 = 0.0051$, $\phidot_0 = -1.414125 \times 10^{-11}$, $\chidot_0 = 0$. See section \ref{Sec:BG_C}. The lower plot shows the scalar ($\mathcal{P}_\mathcal{R}$, blue solid line) and tensor ($\mathcal{P}_\mathcal{T}$, black dashed line) power spectra for this trajectory.}
    \label{fig:BGC}
\end{figure}

Shown in the lower graph of figure \ref{fig:BGC}, the particularly large $\ga$ present in trajectory C leads to a considerable boost in the amplitude of the scalar spectrum and hence a very small tensor-to-scalar ratio. For this kind of set of parameters where $\ga$ is especially large, we find that in the presence of non-constant couplings, $\ga$ typically changes too rapidly during the observational window to produce a spectrum which is close enough to scale-invariance to be observationally feasible. Choosing the constant-coupling limit of our theory, $c = d = 0$, to ensure $\ga$ can remain fairly stable while observable scales are leaving the horizon, we find a spectrum with $A_s = 2.16 \times 10^{-9}$ and $n_s = 0.967$ at the Planck pivot scale, and a tensor-to-scalar ratio of $r_{0.002} = 1.22 \times 10^{-9}$ is produced. The running of the spectral index is $\alpha=-1 \times 10^{-3}$. 

\subsubsection{Trajectory D}\label{Sec:BG_D}
To demonstrate how the evolution of $\gamma$, and the hence the sound speed, affect the perturbation spectra, we have plotted, in figure (\ref{fig:BGD}) a trajectory which is ruled out by observations. As can be seen, the sound speeds vary significantly during the time the observable scales leave the horizon. While the resulting tensor spectra don't show significant running of the spectral index, the scalar power spectra show a rather large spectral tilt as well as significant running of the spectral index. The spectral running is of order $10^{-2}$. 
These features are too pronounced in the plotted example model to be compatible with the Planck data  \cite{Ade:2015lrj}, which limits the running to $\mathcal{O}(10^{-2})$. Analytically speaking, the dependence of the running on rapid changes in $\gamma$ is not unexpected. If we first note that, in single-field inflation, the power spectrum is inversely proportional to $c_s$, we can qualitatively see that since $c_s^{\chi} = 1 / \gamma$, and so loosely $\mathcal{P}_\mathcal{R} \propto \gamma$. This in turn implies that the running, $\alpha$, is affected by changes in $\gamma$ such that,

\beq
\alpha \propto \frac{\bd^2 \ln \gamma}{\bd \ln k ^2} \, .
\eeq

In reality this effect is further complicated by the contribution from the second field's sound speed, $c_s^{\theta}$, which has more complicated dependence on the disformal coupling parameters. Nonetheless, we can say both from this and the numerical verifications carried out that too rapid a change in disformal coupling strength will be constrained by the Planck limits on running of the scalar spectral index.

\begin{figure}
    \centering
    \includegraphics[width=\textwidth]{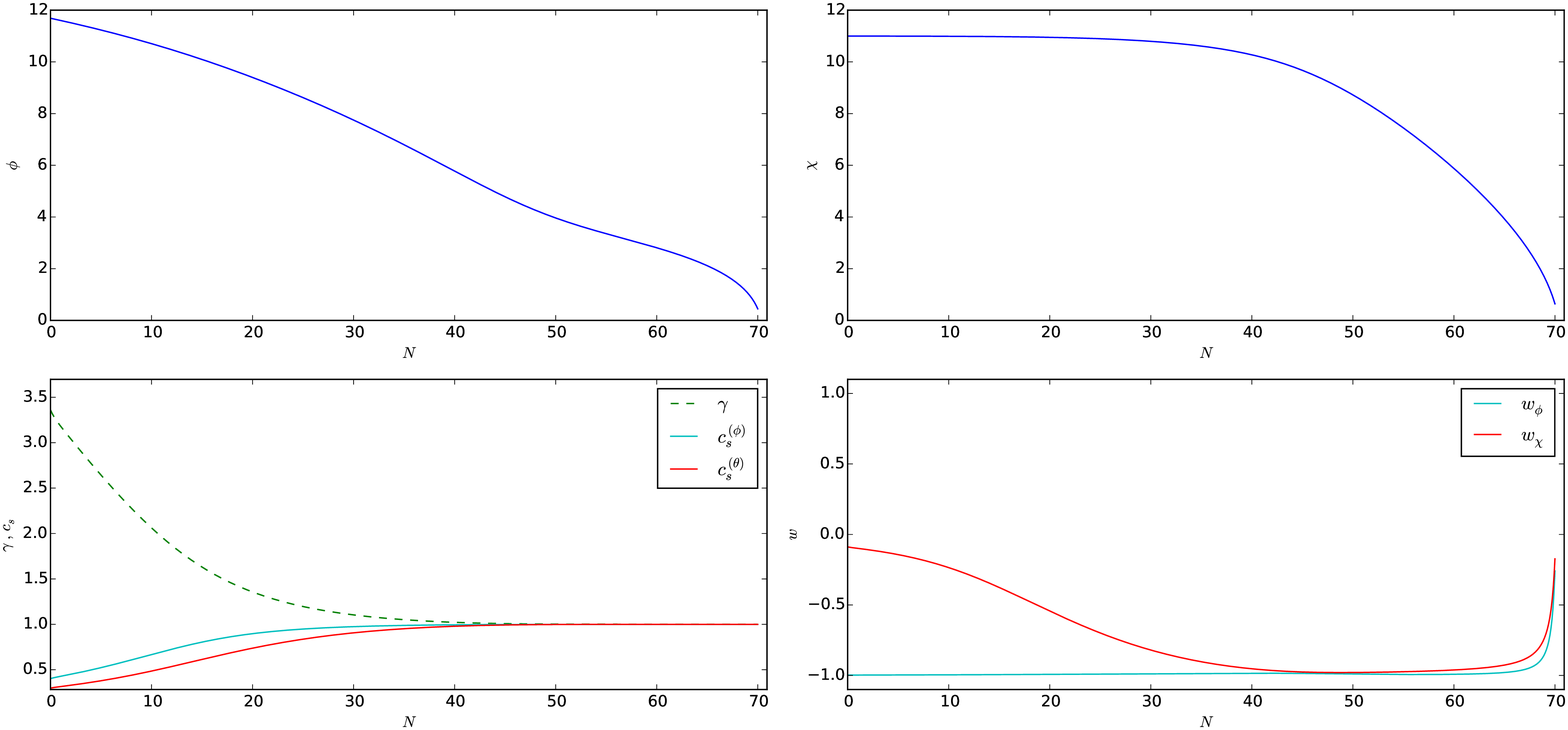}
    \includegraphics[width=\textwidth]{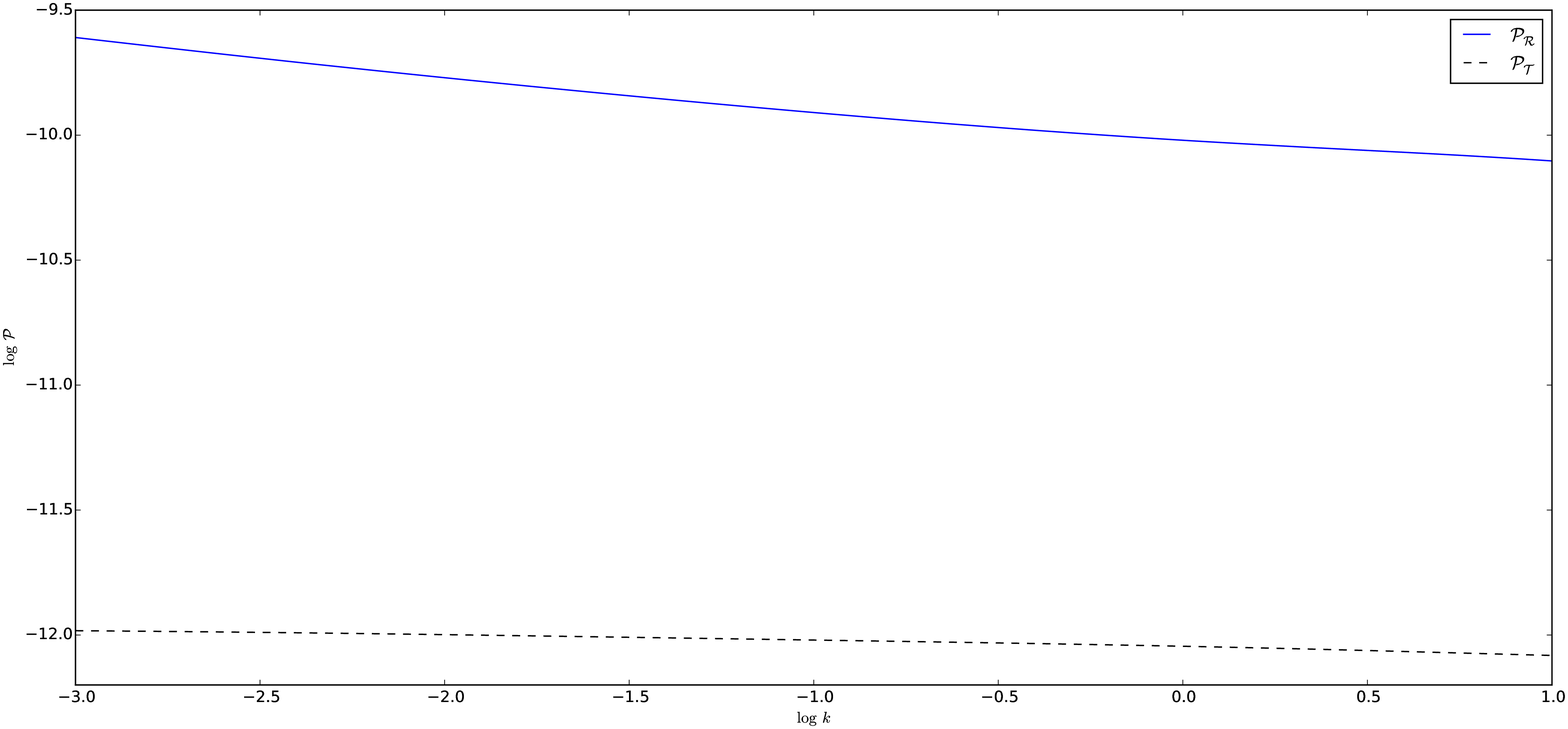}
    \caption{Trajectory D's background evolution of fields, sound speeds and equations of state as a function of the e-fold number $N$ are shown in the four upper graphs. Parameters and initial conditions used are: $d = 0.3$, $c = -0.3$, $D_0 = 5 \times 10^{9}$, $C_0 = 1$, $m_\phi = m_\chi = 1 \times 10^{-6}$, $\phi_0 = 11.689$, $\chi_0 = 11.0$, $\phidot_0 = -4.05\times 10^{-7}$, $\chidot_0 = 0$. See section \ref{Sec:BG_D}. The lower plot shows the scalar ($\mathcal{P}_\mathcal{R}$, blue solid line) and tensor ($\mathcal{P}_\mathcal{T}$, black dashed line) power spectra for this trajectory.}
    \label{fig:BGD}
\end{figure}

\subsubsection{Varying the disformal coupling}
Finally, we show the effect of varying the disformal coupling on the power spectra produced. To be concrete, we choose examples of trajectory $B$. The results for the scalar and tensor spectra are shown in figure \ref{fig:B_change}. As expected, the disformal coupling affects the amplitude of the scalar power spectrum as well as the spectral index (see caption of figure \ref{fig:B_change} for details). For this type of trajectory we observe that for larger $D_0$ the scalar amplitude is larger, while the spectral index is closer to one. The tensor-to-scalar ratio, on the other hand, gets larger for smaller $D_0$.

\begin{figure}
    \centering
    \includegraphics[width=\textwidth]{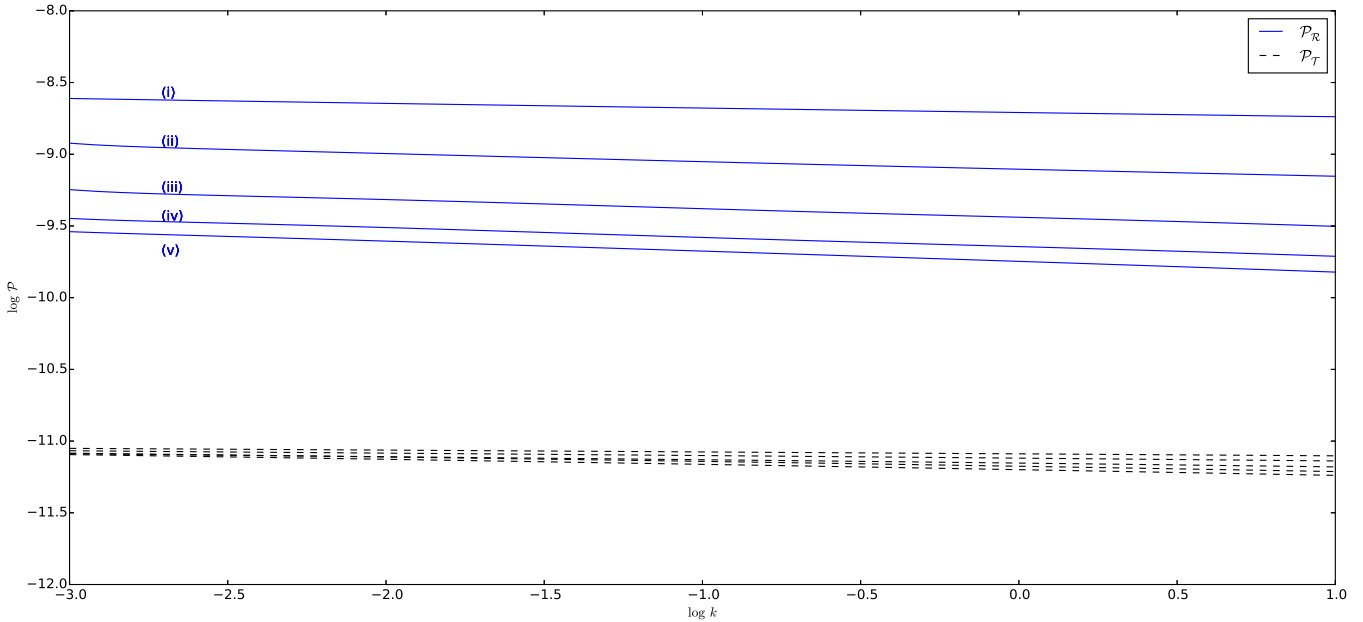}
    \caption{Power spectra for trajectory B with different choices for the parameter $D_0$. The blue curves show the scalar power spectra, the black (dashed) lines show the resulting tensor power spectra. The choices are (i) $D_0 = 8\times 10^9$,  (ii) $D_0 = 4\times 10^9$, (iii) $D_0 = 2\times 10^9$, (iv) $D_0 = 1\times 10^9$ and (v) $D_0 = 0.5\times 10^9$. The spectral indices for these choices of parameter are 0.9677, 0.9435, 0.9385, 0.9346 and 0.9306, respectively. The tensor-to-scalar ratio varies from 0.03 for $D_0 = 8\times 10^9$ to 0.25 for $D_0 = 0.5\times 10^9$.}
    \label{fig:B_change}
\end{figure}

\section{Conclusions} \label{Sec:Conclusions}

In this work we have derived the equations of motion for a scalar-tensor theory coupled to matter propagating on a metric which is disformally related to the gravitational metric. We first investigated a realisation of this model with a power-law coupling motivated by a DBI-type string geometry, and numerically found that for this case, the disformal coupling has little to no effect in the context of inflation, and that this is not entirely unexpected. We then went on to study the model with a phenomenological choice of couplings which were able to produce non-trivial disformal behaviour during inflation, and found that through modifications to the sound speeds of the matter fields and new terms in the equations of motion, disformal effects have a strong influence on the amplitude and tilt of the power spectrum. In particular, the presence of a large $\ga$ during the observable window amplifies the scalar perturbation spectrum with respect to an ordinary two-field model of inflation and hence predicts a low tensor-to-scalar ratio. A special case of the phenomenological model, where the coupling functions are constant, was also found to be able to produce significant disformal effects. \\

We observed that typically, for observationally realistic parameters, the disformal effects vanish at late times and therefore do not lead to significant modifications to the reheating dynamics. This is with the exception of trajectory C (section \ref{Sec:BG_C}), where the large and constant disformal coupling is still present at late times and would seem to produce oscillations in the sound speeds of the matter fields as $\phi$ oscillates around its minimum during post-inflationary reheating. \\

The two scalar degrees of freedom are both stable and causal.
In the limit $\gamma \gg 1$, both the sound speeds (\ref{eq:csphi}) and (\ref{eq:cstheta}) are suppressed by $1/\gamma$. In the braneworld picture this is seen as a stretching of the time lapse for the propagation of fluctuations in a field upon the ultra-relativistically warped brane. A nontrivial evolution of the sound speeds is expected to produce non-negligible non-Gaussianities, which could be worthwhile to determine in further studies of disformally coupled inflation. Besides inflation, the property of the disformal coupling suppressing the scalar sound speed could also be interesting in view of the late-time cosmological dark D-brane scenario. In particular, there the DBI field becomes effectively vacuum energy in the limit $\gamma \gg 1$, while the scalar field upon the brane becomes, due to its equation of state (\ref{eq:wchi}) being suppressed by $1/\gamma^2$, effectively cold dark matter, thereby realising with apparent success, both at the level of background and perturbations, scalar field DIMP matter. 

\acknowledgments{The work of CvdB is supported by the Lancaster- Manchester-Sheffield Consortium for Fundamental Physics under STFC Grant No. ST/L000520/1. CL is supported by a STFC studentship.}

\appendix

\section{Full expressions for the $X_n$, $Y_n$ and $Z_n$ coefficients}  \label{App:XYZ}
\begin{align*}
X_1 =  & \ 2U - \gbm{2}{3} \rhochi - \ga^4 \pchi \, , \\
X_2 =  & \  U' - \frac{1}{2}\left( \left[ \gbm{2}{5}\rhochi + \ga^4 \pchi\right] \cpoc - \gbm{}{1}\left[2 \rhochi + \ga^2 \pchi \right] \dpod \right) \, , \\
X_3 =  & \  \left[1 + \doverc \ga^2 \left(2\rhochi + \ga^2 \pchi\right) \right]\phidot\,, \\
X_4 =  & \  \ga C^2 V'\,, \\
X_5 =  & \ \ga^3 C \chidot\,.  \\
\\
Y_1 =  & \  -\left[1 + \doverc \rhochi \right] \phidot\,, \\
Y_2 =  & \  - \ga C \chidot\,. \\
\\
Z_1 =  & \  -2\left(\phidot^2 - U\right) - \left(\rhochi + 3\pchi \right)\,, \\
Z_2 =  & \ -U' - \frac{1}{2} \left[ \left(\rhochi - 3\pchi\right) \cpoc - \frac{\ga^2 - 1}{\ga^2} \rhochi \dpod \right]\,,\\
Z_3 =  & \  \left[1 + \doverc \rhochi \right] \phidot\,, \\
Z_4 =  & \  -\frac{C^2 V'}{\ga}\,, \\
Z_5 =  & \  \ga C \chidot\,.
\end{align*}

\section{Full expressions for the $\al{n}$ and $\be{n}$ coefficients} \label{App:AlphaBeta}
\begin{align*}
\al{1}  = & \  1 + \doverc \ga^2 \rhochi \, , \quad \al{2}  = \ 0 \, , \\
\al{3}  = & \  -\left(1 - \doverc \ga^2 \pchi\right) \, , \quad \al{4}  = \   0 \,  , \\
\al{5}  = & \  - \phidot \left[4 + \doverc \ga^2 \left(\rhochi - 3 \pchi\right)\right] \, , \\
\al{6}  = & \  3 H \left[1 - \doverc \left(\ga^4 \pchi + \gbm{}{1}\left(\rhochi + \ga^2 \pchi\right)\right)\right] + \doverc \ga^2 \phidot \left[ \doverc \ga^2 \left(4 \rhochi + \ga^2 \pchi \right) \phidotdot \right.  \\
 & \left.  - \frac{1}{2}\left( \left[\gbm{4}{1} \rhochi + \gbp{}{4}\ga^2 \pchi\right] \cpoc - \left[\gbm{4}{2}\rhochi + \gbm{}{1}\ga^2 \pchi\right] \dpod \right) \right] \ , \\
\al{7} = & \  D \ga^3 \left(\ga^2 \phidotdot - 3 H \phidot \right)\chidot - \frac{1}{2} C \ga^3 \left(\gbp{}{1} \cpoc - \gbm{}{1} \dpod \right)\chidot \, , \\
\al{8} = & \  -\left(2 + \doverc \ga^2 \left[\gbm{4}{1} \rhochi + \ga^4 \pchi\right]\right)\phidotdot - 3 H \phidot \left(2 - \doverc \ga^2\left[\rhochi + \gbm{2}{1} \pchi\right] \right) \\
& + \frac{1}{2}\left( \left[ \left(4\ga^4 - 4\ga^2 + 2\right) \rhochi + \left(\ga^4 + 4\ga^2 - 3\right) \ga^2 \pchi\right] \cpoc - \left[\left(4\ga^4 - 5\ga^2 + 1\right) \rhochi + \gbm{}{1} \ga^4 \pchi\right] \dpod \right) \, , \\
\al{9} =  & \ U'' + \frac{1}{2}\left(\left[\gbm{}{2}\rhochi + 3\ga^2 \pchi\right] \dppod - \left[\gbm{}{1}\rhochi\right] \cppoc \right) + \frac{1}{4}\left(\left[\frac{1}{2}\gbm{4}{3}\cpoc - 2\gbm{}{1}\dpod\right]^2 \rhochi \right. \\
& \left. + \left[\gbp{}{2}\cpoc - \gbm{}{1}\dpod \right]^2 \ga^2 \pchi + \left[\frac{15}{4}\rhochi - 13 \ga^2 \pchi \right] \left(\cpoc\right)^2 \right) + \frac{\ga^2 D}{2 C} \left[\left( \left[\gbm{4}{2}\phidotdot - 3H\doverc \phidot^3\right] \rhochi \right. \right. \\
& \left. \left. + \left[\gbm{}{1}\phidotdot - 6H\phidot\right] \ga^2 \pchi\right) \dpod - \left(\left[\gbm{4}{5}\phidotdot - 3H\phidot\right] \rhochi + \left[\ga^4 \phidotdot - 3H\phidot\gbm{2}{3}\right]\pchi\right) \cpoc\right] \, , \\
\al{10} =  & \ \left( \frac{1}{2}\left[\gbm{}{1}\dpod - \gbm{}{5} \cpoc \right] + \doverc\left[\ga^2 \phidotdot + 3H \phidot\right] \right) \ga C^2 V' \,. \\
\\
\be{1} =  & \ \ga^2 \doverc \phidot \chidot \, ,\quad \be{2} =  \ 1 \, , \\
\be{3} =  & \ -\doverc \phidot \chidot \, ,\quad \be{4} =  \ -\frac{1}{\ga^2} \, , \\
\be{5} =  & \ -\gbp{}{3}\chidot \, , \\
\be{6} =  & \ \ga^2 \gbm{2}{1}\doverc\chidot\phidotdot - 2 D \phidot V' - \frac{1}{2}\left[\left(2\ga^4 - \ga^2 - 3\right) \cpoc - \left(2\ga^4 - \ga^2 - 1\right) \dpod \right] \chidot \, , \\
\be{7} =  & \ 3H + \ga^2 \doverc \phidot \phidotdot - \frac{1}{2}\left[\gbm{}{3}\cpoc - \gbm{}{1}\dpod\right]\phidot \, , \\
\be{8} =  & \ -2\ga^2 \left( \ga^2 \doverc \phidot \chidot \phidotdot - C V' - \frac{1}{2}\left[\cpoc - \dpod\right]\gbm{}{1}\phidot\chidot\right) \, , \\
\be{9} =  & \ \ga^4 \doverc \phidot \chidot \phidotdot \left(\dpod-\cpoc\right) + \frac{CV'}{\ga^2} \left[\ga^2 \cpoc - \gbm{}{1}\dpod\right] +\frac{1}{2}\left[\gbm{}{1}\dppod - \gbm{}{3}\cppoc \right. \\
& \left. + \left(\ga^2 \cpoc - \gbm{}{1} \dpod \right)^2 - 3\left(\cpoc\right)^2\right]\phidot\chidot \, ,\\
\be{10} =  & \ \frac{CV''}{\ga^2} \, .
\end{align*}

\section{Full expressions for the $\alb{n}$ and $\beb{n}$ coefficients} \label{App:AlphaBetaBar}

\begin{align*}
\alb{6} =  & \ \al{6} - \frac{1}{2H}\left[\left(\al{1} \phidot + \al{2} \chidot \right) Z_3 + \left( \al{3} \phidot + \al{4} \chidot  \right) X_3 \right]\,, \\
\alb{7} =  & \ \al{7}  - \frac{1}{2H}\left[\left(\al{1} \phidot + \al{2} \chidot \right) Z_5 + \left( \al{3} \phidot + \al{4} \chidot  \right) X_5 \right]\,, \\
\alb{9} =  & \ \al{9} + \frac{Y_1}{2H}\left(\alb{6} \phidot + \alb{7} \chidot + 2\left[ \al{1} \left(\phidotdot - \frac{\dot{H}\phidot}{H}\right) + \al{2} \left(\chidotdot - \frac{\dot{H}\chidot}{H}\right) \right] - H\left[ \al{5} + \left(4 \al{1} - 3 \al{3} \right) \phidot + \left(4 \al{2} - 3 \al{4} \right) \chidot \right] \right) \\ & - \frac{Z_2}{2H}\left(\al{1} \phidot + \al{2} \chidot \right) - \frac{X_2}{2H}\left(\al{3} \phidot + \al{4} \chidot \right)\,, \\
\alb{10} =  & \ \al{10} + \frac{Y_2}{2H}\left(\alb{6} \phidot + \alb{7} \chidot + 2\left[ \al{1} \left(\phidotdot - \frac{\dot{H}\phidot}{H}\right) + \al{2} \left(\chidotdot - \frac{\dot{H}\chidot}{H}\right) \right] - H\left[ \al{5} + \left(4 \al{1} - 3 \al{3} \right) \phidot + \left(4 \al{2} - 3 \al{4} \right) \chidot \right] \right) \\ & - \frac{Z_4}{2H}\left(\al{1} \phidot + \al{2} \chidot \right) - \frac{X_4}{2H}\left(\al{3} \phidot + \al{4} \chidot \right)\,. \\
\\
\beb{6} =  & \ \be{6} - \frac{1}{2H}\left[\left(\be{1} \phidot + \be{2} \chidot \right) Z_3 + \left( \be{3} \phidot + \be{4} \chidot  \right) X_3 \right]\,, \\
\beb{7} =  & \ \be{7}  - \frac{1}{2H}\left[\left(\be{1} \phidot + \be{2} \chidot \right) Z_5 + \left( \be{3} \phidot + \be{4} \chidot  \right) X_5 \right]\,, \\
\beb{9} =  & \ \be{9} + \frac{Y_1}{2H}\left(\beb{6} \phidot + \beb{7} \chidot + 2\left[ \be{1} \left(\phidotdot - \frac{\dot{H}\phidot}{H}\right) + \be{2} \left(\chidotdot - \frac{\dot{H}\chidot}{H}\right) \right] - H\left[ \be{5} + \left(4 \be{1} - 3 \be{3} \right) \phidot + \left(4 \be{2} - 3 \be{4} \right) \chidot \right] \right) \\ &  - \frac{Z_2}{2H}\left(\be{1} \phidot + \be{2} \chidot \right) - \frac{X_2}{2H}\left(\be{3} \phidot + \be{4} \chidot \right)\,, \\
\beb{10} =  & \ \be{10} + \frac{Y_2}{2H}\left(\beb{6} \phidot + \beb{7} \chidot + 2\left[ \be{1} \left(\phidotdot - \frac{\dot{H}\phidot}{H}\right) + \be{2} \left(\chidotdot - \frac{\dot{H}\chidot}{H}\right) \right] - H\left[ \be{5} + \left(4 \be{1} - 3 \be{3} \right) \phidot + \left(4 \be{2} - 3 \be{4} \right) \chidot \right] \right) \\ & - \frac{Z_4}{2H}\left(\be{1} \phidot + \be{2} \chidot \right) - \frac{X_4}{2H}\left(\be{3} \phidot + \be{4} \chidot \right)\,. \\
\end{align*}

\bibliography{refs8}

\end{document}